\documentclass[12pt]{article}
\usepackage{graphicx}
\begin{document}
\large
\begin{center}
\bf { Kepler's Differential Equations}\\[2.0ex]

\vspace*{1.0cm}

  Martin Holder \\ [2.0ex]

\vspace*{1.0cm}

\normalsize
  University of Siegen,
 Fachbereich Physik,
D 57068 Siegen,
 Germany \\[2.0ex]
\end{center}
\vspace{1.0cm}
\begin {abstract}
Although the differential calculus was invented by Newton, Kepler
established his famous laws 70 years earlier by using the same idea,
namely to find a path in a nonuniform field of force by small steps.
It is generally not known that Kepler demonstrated the elliptic orbit
to be composed of intelligeable differential pieces, in modern
language, to result from a differential equation. Kepler was first to 
attribute planetary orbits to a force from the sun, rather than giving
them a predetermined geometric shape. Even though neither the force
was known nor its relation to motion, he could determine the
differential equations of motion from observation. This is one of the
most important achievements in the history of physics.\\
 In contrast to Newton´s {\it Principia} and Galilei´s {\it Dialogo}
 Kepler's text is not easy to read, for various reasons. Therefore, 
in the present article,
his results --- most of them well known --- are first presented in modern
language. Then, in order to justify the claim, the full text of some
relevant chapters of {\it Astronomia Nova} is presented. The translation
from latin is by the present author, with apologies for its
shortcomings. 
\end {abstract} 

\normalsize
\vspace*{1.0cm}\noindent
    {\bf1. Introduction}\\[3.0ex]
 The three laws of Kepler are well known:\\
\begin{enumerate}
\item The orbits of the planets are ellipses with the sun in one of the 
focal points.
\item The radiusvector of the planet covers equal areas in equal times.
\item The squares of the revolution times are in the same relation as
the cubes of the average distances.
\end{enumerate}
Every student of physics learns also how to derive these laws from Newtons
equation of motion and the gravitational law.\par
Less known is how Kepler discovered these laws, except that he found them
when analysing Tycho de Brahes observations of Mars. Almost unknown is that
Kepler set up a complete system of differential equations of motion, i.e.
he indicated the differential change of the planets coordinates r and $\phi$
as a function of time. This he did even for two systems of polar coordinates,
one with the sun at the origin, the other with the center of the ellipse at
the origin. Moreover, he insisted that the latter system which is easier to
integrate cannot be the ultimate solution because the center of the ellipse
as an imaginary point is not accessible to the planet. The planet has to find
its way by looking only to the sun and to the stars. The elliptical orbit is
not predefined but results from the composition of differential pieces.\par
 How important this concept is for the development of science will be clear
to every physicist. For example, Einstein writes in an article about Newton
~\cite{Newton}
:\\{\it
Kepler's laws give a complete answer to the question of how the planets move
around the sun: elliptical form of the orbits, equal areas in equal times,
 relation between principal axis and revolution time. These rules however
do not satisfy our need for causality. They are logically independent rules
without any inner connection. The third law cannot be easily transferred to
a central body different from the the sun ( there is for example no relation
between the revolution times of a planet around the sun and of a moon around
its planet). Most important however, these laws refer to the motion as a whole
and not to the question of how a state of motion is changed from its actual
value to the one immediately following in time. They are, in our present
language, integral laws and not differential laws.\par
  The differential law is the only form which satisfies the need for causality
of a modern physicist. The clear conception of the differential law is one of
the most important mental achievements of Newton. Not only the idea was
required but also a mathematical formalism which, though existing in
rudimentary form, had to be given a systematic shape. Newton found this in
the form of the differential and integral calculus. The question of whether
or not Leibniz independently of Newton invented the same mathematical methods
may be left aside.}\par
  It is the purpose of the present paper to make Kepler's contribution more 
clear, 400 years after the publication of
{\it Astronomia Nova}~\cite{Kepler}, with the hope
 that the Einsteins of future generations will be better informed.
 Needless to say that Einstein himself was a great admirer of Kepler, as
shown at many instances, e.g. in his article for the `` Frankfurter 
Zeitung ~\cite{Einstein}
on the occasion of the 300th anniversary of Kepler's
 death. He would have admired Kepler even more if he had known the 
following chapters of Kepler's {\it Astronomia Nova}.\par
  When dealing with an issue that is several hundred years old, one might of
course ask, why was there no earlier notice ? Probably there are several
reasons. One of them is language. Not only at present hardly any scientist
is familiar with Latin, also the mathematical language has evolved
enormously since Kepler's time. In {\it Astronomia Nova} there is no single
equation, because the equal sign (=) was not yet invented, there is no
abbreviation with letters, like r for radius. Everything has to be 
explained by words. The only help comes from geometrical drawings. A second
problem is, that Kepler had a wrong notion of forces, inherited from 
Aristoteles. According to Aristoteles the velocity of an object is
proportional to the force applied to it and has the same direction. As is
well known, it was Galilei who first questioned this opinion, saying that
a piece of lead of 100 pounds does not fall from a height of 100 cubits 
\footnote{about 45m ( note by the editor)}
ten times as fast as a piece of 10 pounds. It was him who established by
measurements the true relation between force and motion. His results, among 
them a very elegant, but mathematically incomplete argument about the
centrifugal force, were not published until 1632, two years after Kepler's 
death. The centrifugal force was not known to Kepler, neither the word nor 
the fact. The correct description of it is due to Huyghens (in 1673).\par
  It is understandable that after Newton hardly anybody should have taken
the pain to labour through Kepler's  {\it Astronomia Nova}. In this text,
Kepler's ideas about the laws of motion are mixed with his
notion on forces. To separate them would unavoidably
alter the text. Therefore the full text of the most relevant
chapters including side remarks will be given below, with no attempt to
condense it or even to change the sequence of words, if not necessary.
 It is not always easy
to translate Kepler's text into a modern language, partly because the concepts
have changed. This is especially true for the concept of forces. Apparently
Kepler distinguishes forces applying to dead matter ({\it vis naturalis})
and forces applying to living matter ({\it vis animalis}). The latter is
translated here as `` animated force ``. Examples are given in Kepler's
text.\par
  Readers who want to skip the speculations about forces can easily do so;
they will have no problem to understand the subsequent arguments for 
differential equations. On the other hand, Kepler's text is a fascinating
document in itself. The fact that
he published his ideas, even though he knew that they cannot be correct,
probably reflects the same spirit as advocated in our times e.g. by
Feynman, that theoreticians should not only talk about their achievements,
but also indicate where they failed. The subsequent solution of the puzzle
is then much more instructive.\par
 It will not be difficult for a modern reader to condense the text, after
 having read it, into a more suitable form.\par
  In a certain sense, Kepler's differential equations are not just an
attribute  to an already existing beautiful theory, but they are a logical
necessity right from the start. Kepler's initial concept was, that planet
coordinates should be referred to the sun, because it is the sun, the heart
of the world, as he says, which turns the planets around. His first
occupation with Tycho was to convert a table of Mars oppositions from the
traditional reference to the center of the earths orbit into a table of
Mars oppositions to the sun. Tycho saw that enterprise with a certain
suspicion but he let Kepler do, because he and his longtime assistant
Longomontanus had finished the Mars theory --- even though not everything
fitted exactly as they wished --- and had started to work on the much more
complicated motion of the moon.\par
  Whenever a body moves in a field, where strength and direction of the force
change continuously due to its very motion, then at a given moment only the
immediately following piece of its trajectory can be predicted. Therefore
already in the concept of the sun as a source of forces, a planets orbit
as a solution of differential equations is inherent. That the elliptical
orbit, in contrast to an eccentric circle,
 has in fact this property, to be the solution of a differential
equation, must have given Kepler enormous satisfaction and additional
confidence in the correctness of his result.\par\newpage
  It is somewhat unfortunate and also unfair, that the historical reception
has reduced Kepler's work to the famous three laws. In reality the essential
step from the traditional way of a purely geometric description of
planetary orbits to a physical description was his idea of a force which
comes from the sun. It is this concept which leads to differential equations.
It may be compared to the introduction of the idea of
a field instead of forces which act
over long distances at the end of the 19th century. As the reader will see,
the application of the idea was not immediate; Kepler started in the
traditional way with a geometric concept. As Einstein phrased it : with one
foot he was still in the medieval age. This is well known from Kepler's
biography, but it shows also up in his astronomical work. Yet it was him who
made the first, essential step. The later transition to infinitesimally small
increments looks from this point of view rather like a straightforward
refinement. In our teaching there ought to be a corresponding change.\\[2.0ex]

\noindent
{\bf2. Kepler's differential equations in modern form}\\[2.0ex]

Let a planet P (Fig.1) 
\begin{figure}[hhh]
\hspace*{2.2cm}
  \includegraphics[scale=0.23]{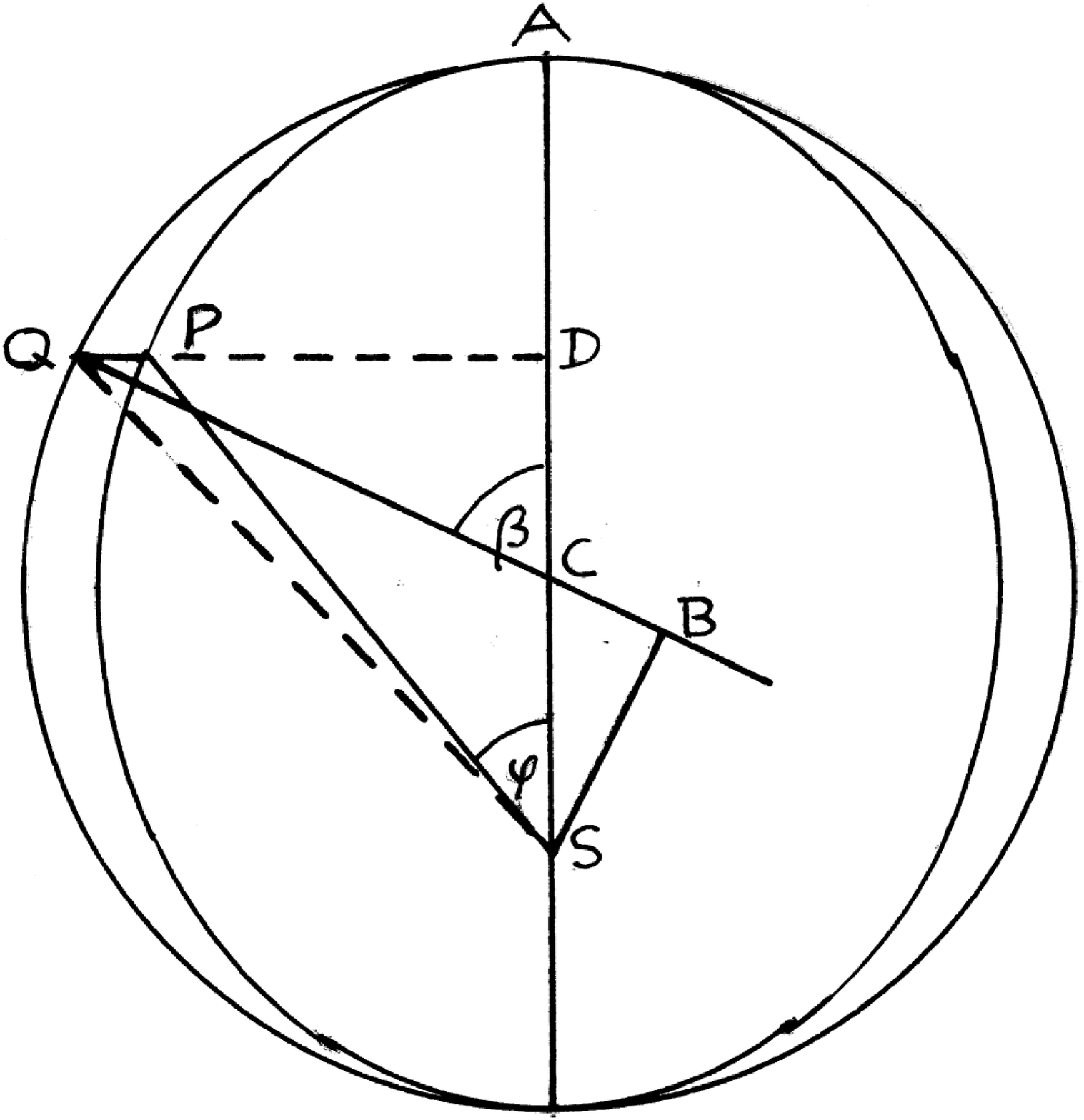}\\
 \caption{
The ellipse with the sun S in one of the focal points and a
planet P which appears at an angle $\phi$ to the principle axis. In Kepler's
time the angle $\phi$ was measured from the aphelion A, nowadays it is
measured from the perihelion. ( see text)}\label{fig:ellips1}
\end{figure}
move around the sun S in an elliptical orbit centered
at C. The sun S is located in one of the focal points of the ellipse. The
so called ``true anomaly'' is the angle $\phi$ between the radiusvector PS
and the principle axis of the ellipse ( the ``line of apsides''). The point Q
is the point which corresponds to P on the circle centered on C with
radius CA equal to the principle axis. The angle $\beta$ between the line QC
and the principle axis is called the ``eccentric anomaly'' of the planet.
The perpendicular drawn from S to the line QC or its extension intersects
this line in B. For later reference, we mention here, that the angle SPC
in which the eccentricity CS appears to the planet is referred to as
``optical equation `` by Kepler ( the term ``equation `` in astronomy
denotes an angle, usually the angle between the real sun and a fictitious
sun moving with constant angular velocity, as seen from the planet ).\\
Let e= CS and r=PS, and CA=1.\\
The area F covered by AQS is then the sum of the circular sector AQC and the
triangle QCS
\begin{eqnarray*}
2 F = \beta + e \sin \beta .  
\end{eqnarray*}  
This equals to $\omega t $ according to Kepler's second law, when t is the
time elapsed since the transition of the aphelion ( in A) 
and $\omega = 2\pi/T $, T = time of revolution, 
\begin{eqnarray*}
\omega t = \beta + e \sin \beta .  
\end{eqnarray*}
This is known as Kepler's equation.\\
Because the area covered by the ellipse differs from the area
covered by the circle by the constant amount $\sqrt{1-e^2}$, Kepler's
 second law      is also valid for the ellipse. In differential form it
says
\begin{eqnarray*}
 r^2 d  \phi = \sqrt{1-e^2 }\  \omega \ d t.
\end{eqnarray*}
An important relation is QB=PS =r. From Fig.1 one reads
\begin{eqnarray*}
QB & = & QC +CB\\
 & = & 1+ e \cos \beta .
\end{eqnarray*}
On the other hand, in the triangle PDS one has
\begin{eqnarray*}
(PS)^2 &=& r^2\\
 &=& (DS)^2 +(PD)^2\\
 &=& (e+\cos \beta)^2+(1-e^2) \sin^2 \beta\\
&=& e^2+2 e \cos \beta +1 -e^2 \sin^2 \beta\\
&=& 1+2 e \cos \beta +e^2 \cos^2 \beta\\
&=& (1+e \cos \beta)^2 , \hspace*{1.0cm}q.e.d.
\end{eqnarray*}

\begin{eqnarray}
 From \hspace{1.0cm} r& =& 1+ e \cos \beta    \hspace{1.0cm} follows\\
 dr&=& e\ d(\cos \beta) = -e \sin\beta \ d\ \beta.     
\end{eqnarray}
This is one of the differential equations which, together with Kepler's 
second law, forms a complete system for  r(t), $\beta$(t). Kepler uses the 
integral
\begin{eqnarray*}
 \int_{r_{max}}^rdr = e ( 1- \cos \beta) ,
\end{eqnarray*}
for  $(1-\cos \beta $) he writes {\it sinus versus $\beta$}. Another
expression, in use both at his time and nowadays, is {\it sagitta}, the
piece of an arrow which sticks out over the chord ( Fig. 2).\\
\begin{figure}[hhh]
\hspace*{2.2cm}
  \includegraphics[scale=0.23]{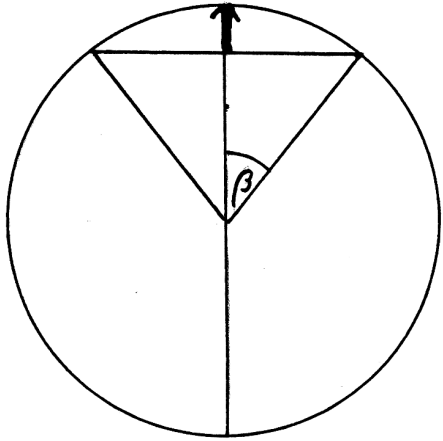}\\
 \caption{
 Geometric illustration of the {\it sinus versus} $\beta$. The
{\it sinus versus} $\beta$ is the sagitta in a circle with unit radius. It
equals ( 1- $\cos  \beta$).}\label{fig:sagitta}
\end{figure}\\[1.0ex]
To obtain polar coordinates centered on the sun we read from Fig.1
\begin{eqnarray*}
 r \cos \phi = e + \cos \beta .
\end{eqnarray*}
Replacing $\cos \beta$ from equation (1) gives
\begin{eqnarray*}
 r \cos \phi& =& e + \frac{r-1}{e}\\
e\ r \cos \phi &=& e^2 +r-1\\
r\ (1-e \cos \phi) &=& 1-e^2\\
\frac{1-e^2}{r}&=& 1- e \cos \phi , \\
\end{eqnarray*}
or differentially
\begin{eqnarray*}
(1-e^2) d(\frac{1}{r}) &=& -e \ d(\cos \phi)\\
d(\frac{1}{r}) &=& - \frac{e}{1-e^2} \ d(\cos \phi)
\; \propto \; \sin \phi \ d\ \phi\\
\end{eqnarray*}
This is the relation referred to by Kepler when he says, $\sin \phi$ is the
measure for the change of the apparent diameter of the sun, i.e. the angle
under which the sun appears from the distance~r. It is the second differential
equation which, together with the second law, completes the system r(t), 
$\phi$(t). The general integral of this equation is
\begin{eqnarray*}
 \frac{1}{r} = a +b \cos \phi ,
\end{eqnarray*}
with arbitrary constants a and b , the equation of a cone section.\\[3.0ex]
\begin{figure}
\hspace*{2.2cm}
  \includegraphics[scale=0.23]{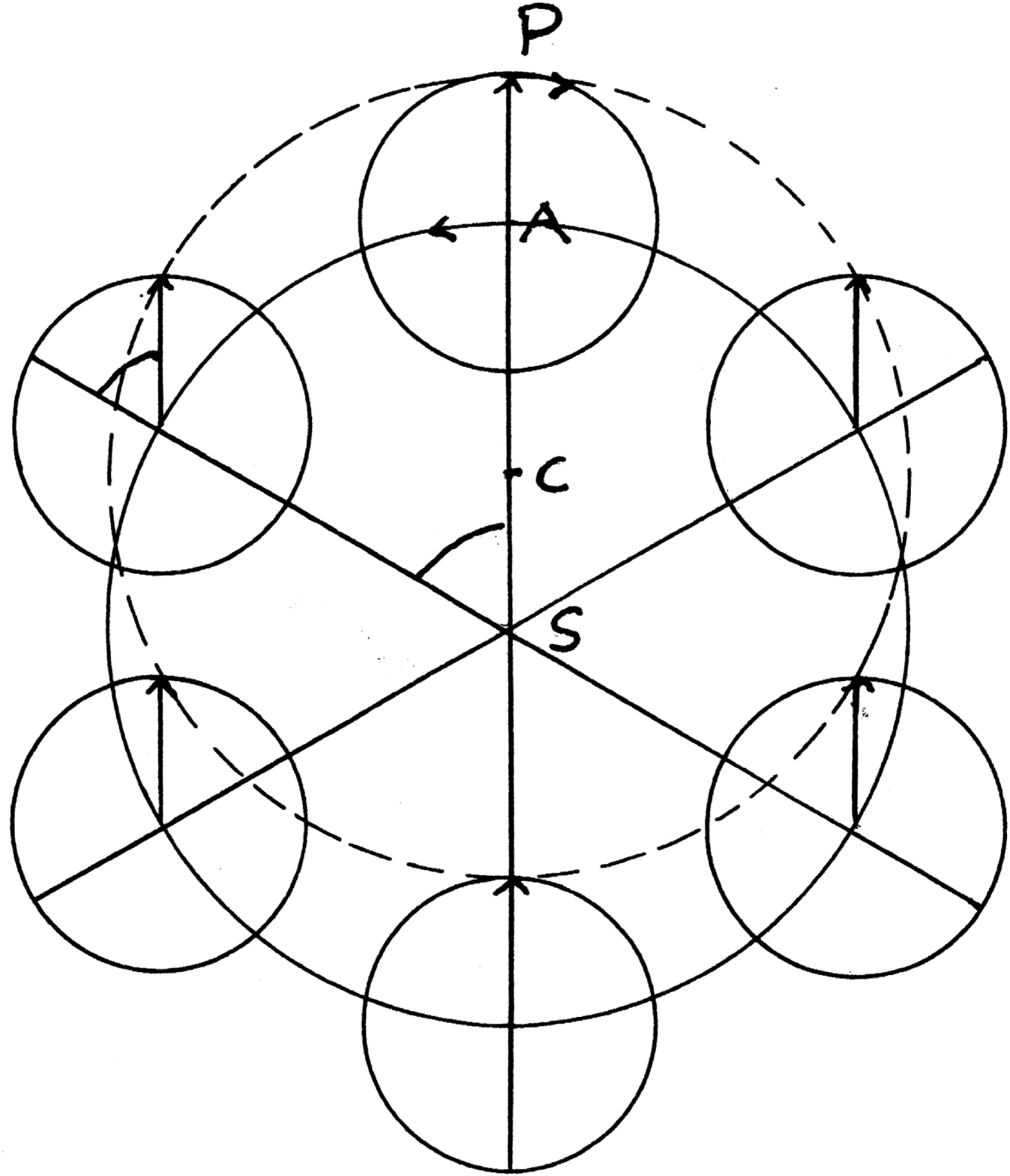}\\
 \caption{ Construction of an eccentric circle ( dashed line) on which a
planet P turns around the sun in S, by the superposition of two circles, one
with large radius SA around the center S, and a second one ( the epicycle)
with smaller radius ( equal to the eccentricity CS) around the tip of the
radiusvector of the former circle \mbox{( see text)}( from {\it Astronomia
  Nova})}\label{fig:epicycle}
\end{figure}\\[1.0ex]

\noindent
{\bf3. Kepler's oval hypthesis as a precursor to the ellipse.}\\[3.0ex]
  In order to understand the subsequent sections it is necessary to
introduce Kepler's idea how to modify a circular motion. He starts from
a construction due to Appolonius, by which the apparent nonuniform motion
of the sun can be explained as a superposition of two uniform circular motions
( Fig.3).
 In the later heliocentric system , a planet  P is mounted on a small
circle, the epicycle, the center A of which moves uniformly around a center S,
the sun. If the motion of the planet on the epicycle is in opposite direction
to the motion of the large circle, say clockwise, if the large circle moves
anticlockwise, and both have the same time of revolution, the planet stays
always in the same direction from the center of the epicycle ( upwards in
\mbox{Fig.3 )}, and so performs a circular motion , the center C of which is
now displaced from the sun S by an amount equal to the radius of the epicycle
( the epicycle is here thought to be fixed like on a spoke to the large 
wheel). Kepler assumes that the reader understands this concept from the
drawing. This is the traditional way how the motion of the earth around the
sun ( or in antiquity, the sun around the earth) was explained. The orbit of
the planet is a circle, called the eccenter.\par
  Now two observational facts on Mars called for a modification of this
  scheme. One is, that the motion in the eccenter is still not uniform,
the other is, that the orbit is apparently not a perfect circle, but is 
narrower along the line of apsides. Kepler's modification to the scheme of
Appolonius was to give the planet a nonuniform motion, satisfying his second
law, while the motion of the planet on the epicycle remains uniform. Since the
motion on the orbit is slower near the aphelion, where the distance to the
sun is largest, according to the second law, the motion of the epicycle is
advanced in the orbit section following the aphelion. The planet deviates
therefore from the circle to the inside. The lead of the epicycle ``clock''
diminishes as the planet approaches the perihelion; there it vanishes.
In the subsequent section from perihelion to aphelion the lead is changed
to a lack, again causing a deviation to the inside. Everyone will agree
that this is an ingenious idea; there is no single free parameter. The
mechanism of such a system however is totally unclear; it also contradicts
Kepler's own premises ( that a circular motion around an imaginary point
makes no sense ). His excuse is funny enough \mbox{(`` Speedy} dogs have blind 
offsprings `` ). As seen from Fig.4,
\begin{figure}
\hspace*{2.0cm}
  \includegraphics[scale=0.23]{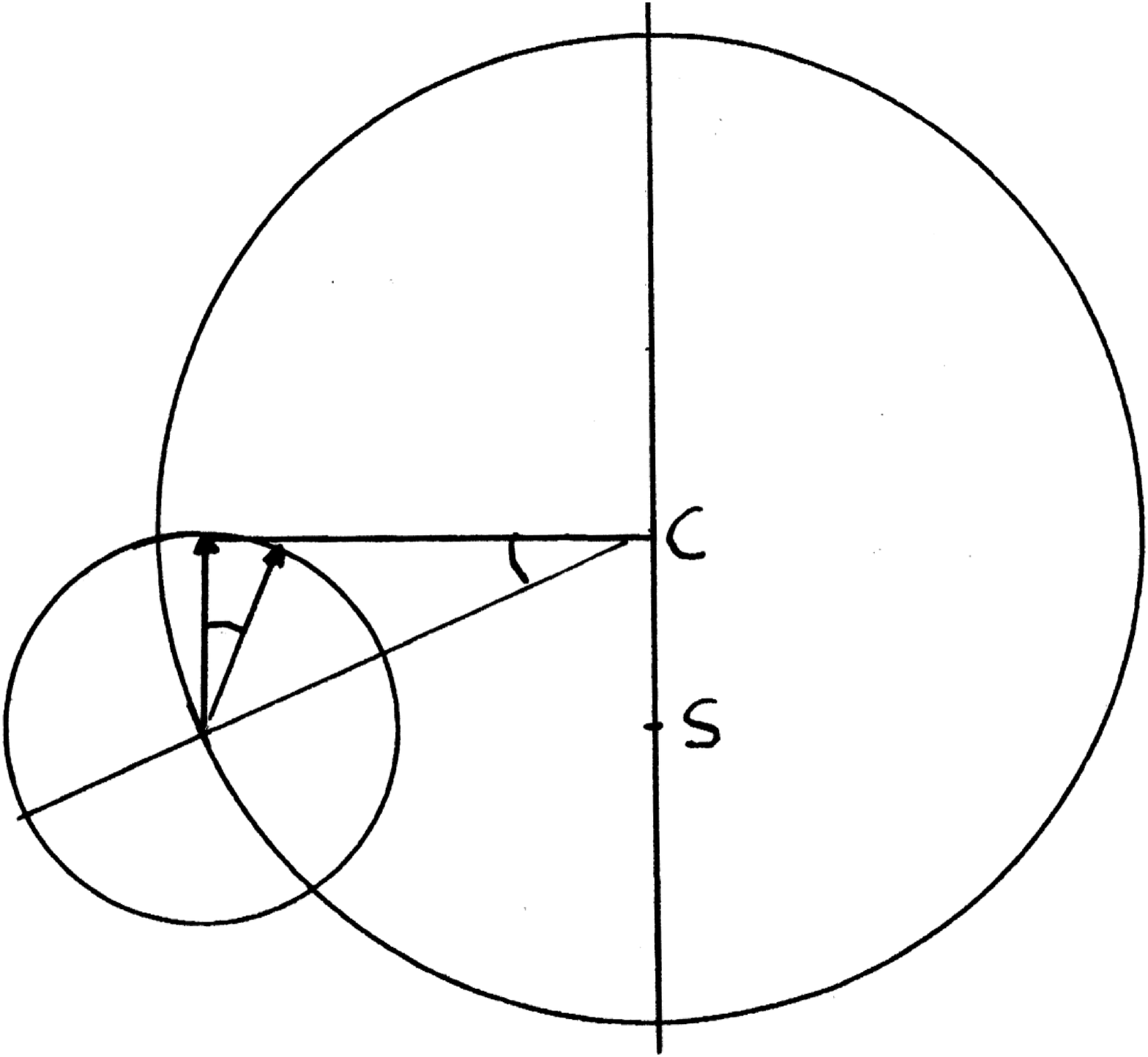}\\
 \caption{ Deviation from a circle in Kepler's oval, at a position of the 
planet at 90 $^0$ to the line of apsides. According to Kepler's second law the
planet has finished a quarter of the total revolution. The center of the 
epicycle in a circular path would be, according to the construction of fig.3,
below the planet, by an amount e. The radiusvector of an epicycle running
at constant speed is, in this position, advanced by an angle as indicated.
Therefore the planet deviates from a circle by an amount e$^2$.
  }\label{fig:gap}
\end{figure}
the deviation from the circle at
\mbox{ 90 $^0$}
from the apsides  is around $e^2$, twice as much as in the final ellipse
($\sqrt{1-e^2} \approx 1-e^2/2$). The difference, $e^2$/2, is three to four
times the observational error, that is, it is significant. There are
inconsistencies also in the time distribution. It is not necessary here to
recall Kepler's efforts to save his oval hypothesis, by modifications of his
second law. This work cost him the entire year 1604, because the numerical
integrations in steps of $1^0$, with five digit accuracy, for three different
eccentricities, had all to be done by hand; logarithms were not yet invented.
The discrepancy at 90 $^0$ , alluded to above, plays a central role in the
discovery of the ellipse, in chapter 56 of the
{\it Astronomia Nova}, 
as will be seen below.\par
Besides a coordinate system fixed in space, Kepler likes to use a system in
which the line joining the sun and the center of the epicycle has a fixed
direction. The transformation between the two systems is, of course, only
simple, if the motion of the epicycle center is known. It is not astonishing
that in the course of the investigation this fictitious point, together with
the whole epicycle, disappears into nothing.\par
\clearpage \noindent
 {\bf4. Kepler's axioms of planetary motion}\\[2.0ex]
 It is significant for Kepler's rigorous thinking that he formulates the 
foundation of his calculations as axioms, as he had already done in his work
  on optics. His axioms are:
\begin{enumerate}
\item The planets tend to rest at the place where they are positioned alone.
\item They are transported by a force which comes from the sun, from place
to place along the ecliptic.
\item If the distance of a planet from the sun is unchanged, the planets
orbit will be a circle.
\item If the same planet would move at different constant distances from
the sun, the times of revolution would be in the ratio of the squares of
the distances.
\item The force which is inherent to the planet is not sufficient to transport
it from one place to another, since the planet has nor legs nor wings nor fins
to support itself on the ether.
\item Inspite of this, the planets approach and regression from the sun is due
to a force inherent in the planet.
\end{enumerate}
All these axioms, says Kepler, are uncontradictory and are in agreement with
nature, according to our present knowledge.\par
  Maybe it is appropriate to add here a few comments before continuing.
The third axiom sounds trivial: a circle is defined as a curve in which every
point has the same distance to the center. The axiom is nontrivial, however, if
one consideres the third dimension. Then the axiom says, the planets orbits are
planes. It was one of Kepler's first actions to convince himself by several
independent methods that the orbit of Mars is in fact , like the orbit of
the earth, a plane around the sun, with a constant inclination to the 
ecliptic. This content of the axiom is however not immediately obvious. Had
Kepler said: the orbits of the planets are planes in which also the sun is
located, he would have said more and it would be clearer. This is an example
of a certain lack of refurbishment which characterizes the whole of
{\it Astronomia Nova}, certainly in part due to various difficulties in the
edition, not to mention Kepler's fragile health.\par
Also the forth axiom calls for a second thought. The situation contemplated
here is somewhat unrealistic. Nobody can place a planet into a different
circular orbit. What the axiom probably wants to say, is, that, if a planet
in the apsides, where the distance from the sun is momentarily constant,
would continue its orbit with the same speed v and at the same distance r,
then its revolution time would be proportional to the square of the distance.
The equality of the product $r*v$ in the apsides is the empirical fact upon
which later Kepler's second law is based. He could have formulated the axiom
as $r_1*v_1 = r_2*v_2$ , where the indices refer to aphelion and perihelion.
That is all he needs. The confusion in the literature about the difference
between the constancy of $r*v$ and of $r*v_\perp$ ( which is the second law)
on the whole orbit, this confusion is in part Kepler's own fault. His own
very sophisticated investigations about this difference on a circular orbit
will not be dealt with here, but they are most likely essential to his 
discovery of the ellipse, as treated in chapter 56 of 
the{\it Astronomia Nova} ( see below).\par
The fifth and sixth axioms concern the forces. As will be seen, Kepler
favours a kind of magnetic force, although he must admit, that the earth's
magnetic field would have the wrong direction. So he must leave the question
 open. One might wish that he had lived long enough to see that the 
gravitational force invented by himself to explain the tides as attraction
by the moon is in fact all what is needed to explain planetary motion.\\[3.0ex]
   {\bf5. Arguments against circular orbits.}\\[2.0ex]
It is historically very interesting to see how the circular orbit,
sacrosanct to the ancient Greeks, disappears in the course of time.
Already Copernicus had given it up, in favour of another principle, the
uniformity of motion. Both principles are wrong. Kepler started from
 the assumption that a force from the sun is responsible for the motion.
It is remarkable that he could find the laws of motion knowing neither the
force nor the relation between force and motion. In chapter 39 of 
{\it Astronomia Nova}, 
after presenting his axioms, he argues that a circular orbit is not compatible
with these axioms and gives a preview of his later discovery of the 
elliptical orbit, leaving aside his own idea of an oval, that failed for
various reasons. Apparently Kepler introduced this chapter 39 in order to
familiarize the reader with the general concept before getting lost in the
details of the calculations.\par
  In the presentation of the text, we will, as for chapters 56 and 57 
below, also include Kepler's comments, presented in the original edition
in small letters on the margin of the page. Here they are inserted in italics
into the text, approximately at the same place as they appear in the original.
Footnotes in the original, characterized by Kepler with a * in the text,
are given as footnotes. A few footnotes by the present editor are 
characterized as such. The translation from Latin is not always 
straightforward. In case of doubt, the reader is asked to consult the original
text.\par
 A short comment will be added at the end of Kepler's text. It may help the
reader in a second reading.\par
 After his presentation of the axioms, Kepler continues in chapter 39:\\[2.0ex]
{\bf 5.1 What the planet achieves by its motion, if the orbit due to the 
composition becomes a circle, i.e. how will the distances from the sun 
be obtained.}\\[1.0ex]
  Let us now exercise with geometrical figures to see which laws are
  necessary to represent any planetary orbit. Let the orbit be a circle as
believed hitherto, eccentric to the sun, the source of the force (
Fig.5).
\begin{figure}[hhh]
\hspace*{2.2cm}
  \includegraphics[scale=0.23]{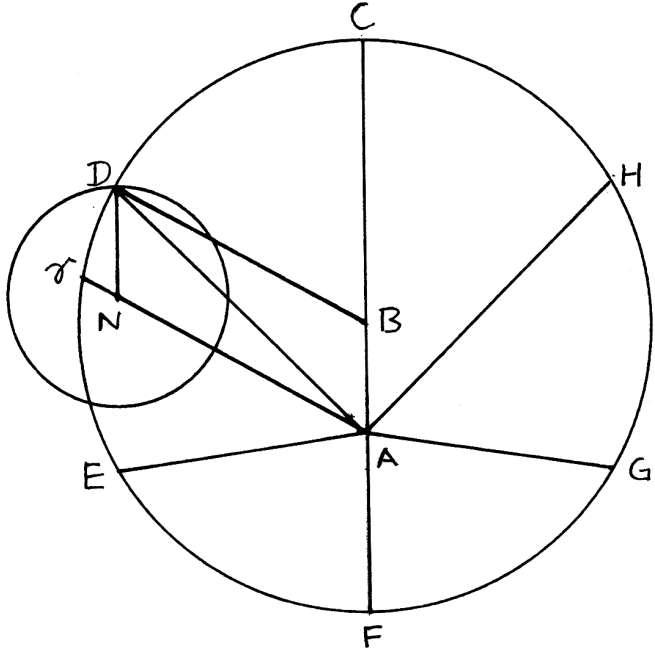}\\
 \caption{ Schematic of an eccentric circular orbit around the sun in A
( from {\it Astronomia Nova}).
  }\label{fig:circle}
\end{figure}
The eccenter shall be CD, with center B, radius BC; the line of apsides shall 
be BC, the sun A and BA the eccentricity. The eccenter is divided into an
arbitrary number of equal parts, starting from C on the line of apsides;
their ends shall be connected to A. Therefore CA,\ DA,\ EA,\ FA,\ GA,\ 
HA will be 
the distances from the source of the end points of equal parts. Also
(Fig.6)
\begin{figure}[hhh]
\hspace*{2.2cm}
  \includegraphics[scale=0.3]{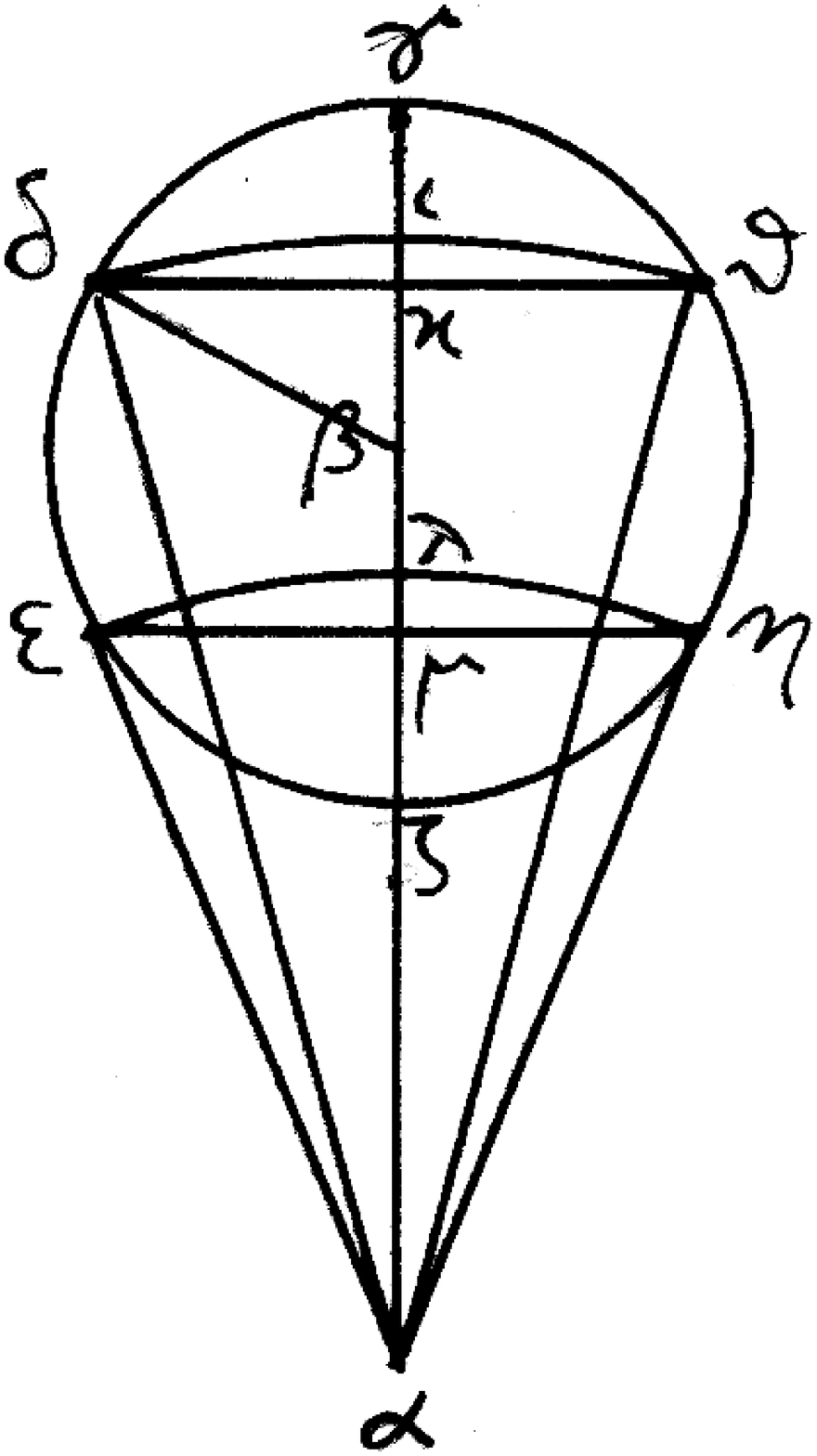}\\
 \caption{Positions of a planet in an epicycle. The sun is in $\alpha$,
the positions on the epicycle correspond to the positions on the circular 
orbit in fig.5 ( from {\it Astronomia Nova}).  }
\label{fig:smic}
\end{figure} 
an epicycle $\gamma \delta $ shall be described around a center $\beta$ 
with radius $\beta \gamma$ equal to AB, and be divided into the same equal
parts as the eccenter, starting from $\gamma$. The line $\beta \gamma $ shall
be continued such that $\beta \alpha $ equals BC, and the point $\alpha$ 
shall be connected to the end points of the equal parts in the epicycle in the
lines $\gamma \alpha,\delta \alpha,\epsilon \alpha,\zeta \alpha, \eta \alpha,
 \theta \alpha$.
These lines will be equal to the distances drawn from A in the eccenter,
respectively. This was already demonstrated in chapter 2 above. Now the arc
$\delta \iota \theta $ shall be descibed around the center $\alpha$ with
radius $\alpha \delta$; it intersects the diameter$ \gamma \zeta$ in
 $\iota$; around the same center with radius $\alpha \epsilon$ the arc
$\epsilon \lambda \eta $ intersects the diameter $\gamma \zeta $ in 
 $\lambda$. 
 The terminal points with equal distances from the aphelion 
$\gamma$ of the epicycle shall be connected in the lines $\delta \theta, 
\epsilon \eta$, which intersect the same diameter in $\kappa, \mu$, such that
$\alpha \delta $ or $ \alpha \iota$ is longer than $\alpha \kappa $, and
$ \alpha \epsilon $ or $ \alpha \lambda $ longer than $\alpha \mu $.\\[1.0ex]
{\it \bf  First mode: the planet itself runs the epicycle.}\\[1.0ex]
 If it were possible that the planet describes a perfect epicycle by an
 inherent force and at the same time its orbit becomes a perfect circle, then
similar arcs would have to be completed simultanously, in the eccenter and
in the epicycle. Therefore, it is immediately obvious, by which means, by
which measure the distances $\iota \alpha $ and AD are made equal. Namely,
since $\alpha \iota $ and $\alpha \theta $ are equal, the planet going from
$\gamma$ to $ \theta$  will necessarily and without special advice find the
right distance $\alpha \theta $ equal to AD.\\
{\it The absurdities of this mode: first absurdity}\\
But besides that it seems to be in conflict with the fifth axiom if one says
that the planet proceeds by an intrinsic force from place to place, there are
many other absurdities involved. \\
Let AN be parallel to BD and AN be equal to BD. The epicycle around the center
N will go through D. Now, if CD is on a perfect circle, the angles of the
planet at the center B of the eccenter and that of the epicycle center N at the
sun A will be equal ( by the equivalence demonstrated in chapter 2), such that
the epicycle diameter ND, on which the planet sits in D, will remain parallel
to AB in space. Therefore the velocities of the epicycle center N around the
sun in A and of the planet D around the center B of the eccenter will be
the same; the motions will simultanously intensify and relax, and, because of
this intensification and relaxation is due to the larger or smaller distance
of the planet to the sun, the center of the epicycle, which remains always\\
{\it Second absurdity}\\
in the same distance, would have to move slowly or fast because of the larger
or shorter distance of the planet from the sun.\\
{\it Third absurdity}\\
And, while the force driving the planets is always faster than the planets,
as shown in chapter 34, we would here have to assume one of the rays from the
sun, AN, i.e. the line on which the center of the epicycle remains, to be\\
{\it Fourth absurdity}\\
sometimes late, sometimes early, again in contrast to what was said before,
that a force in the same constant distance produces the same velocity.\\
 {\it Fifth absurdity}\\
The planet however would have to be assumed to evolve from this imaginary ray
AN in opposite directions unequally in equal times, in order that this ray
itself becomes fast or slow.\\
{\it This last statement will be denied to be absurd below in chapter 49,
while the other absurdities remain }\\
In this way we would be closer to the assumptions of the ancients but deviate
enormously from our physical speculations, as shown in chapter 2. Also, my
imagination is not sufficient to see a way in which this mode could be
realized in nature.\\
{\it \bf Second mode, that the planet moves the eccenter.}\\[1.0ex]
It seems therefore simpler, if we look at the diameter ND of the epicycle
which remains always parallel to itself. The planet will effect its motion
inspecting not the epicycle but the center B of the eccenter and by keeping
always the same distance from this center.\\
 {\it First absurdity}\\
But in the beginning of this opus, in chapter 2, it was said, that it is
totally absurd, that a planet( by whatever intelligence) figures itself a
center and a distance from it, if the center is not marked by a special body.
And also , if somebody says, the planet looks to the sun A and knows
beforehand from memory, which distances from the sun it should keep in order
to stay on a perfect eccenter, then this is even more remote. It also lacks
 the means to connect the perfectly circular orbit with the signs of 
increasing and decreasing diameter of the sun, and this for any whatsoever
intelligence. These means are nothing else than to position the center B of
the eccenter in a certain distance from the sun, what, as we already said,
cannot be accomplished by an intelligence alone.\\
 I do not deny, that one can imagine a center and a circle around it. But this
 I say, if this center exists only in the imagination, without indication of
time or an external sign, then it is impossible to order in reality any
mobile body into a perfectly circular orbit around it.\\
 {\it Second absurdity}\\
In addition, if a planet would take its correct distances required for a
circular orbit, from memory, it would take as well, like from the Prutenic
or Alphonsinic tables, the equal arcs of the eccenter to be run in unequal
times and by the external force of the sun, and so it would prescribe from
memory, what the alien force of the sun should be effecting. All this is
absurd.\\
 {\it Third absurdity}\\ 
Especially since according to Aristoteles there is no science of the infinite;
the infinite however is mixed in this intensification and relaxation.\\
 But luckily, also the observations themselves do not support a perfect
 circle, as will be shown below is chapter 44. So these ( seemingly) foolish
speculations are not alone and are all the less subject to disparaging
remarks.\\[1.0ex]
 {\it \bf Third mode, that the planet librates by an inherent force on the
diameter of the epicycle.}\\[1.0ex]  
It is therefore more likely that the planet itself does not care for the
epicycle nor for the eccenter but the work which it effects, or to which it 
contributes, consists of a libration or balancing movement on the diameter $
\gamma \zeta $ tending towards the sun.\\
By which measure does the planet measure the correct distance at any given
time?\\
{\it The planet cannot obtain the correct distances from a real epicycle}\\
 To us the measure is is apparent from geometry and a drawing. Whenever the
planet is promoted by the solar force to the line AD, we look for the angle
CBD and make $\gamma \beta\delta $ the same. And so we say, $ \alpha \delta $
 or the equal amount $\alpha \iota $ is the correct distance of the planet 
in D from the sun A. But this measure, available to us humans, we have already
taken away from the planet by forcing it from the width of the epicycle into
the narrowness of the diameter $ \gamma \zeta$.\\
{\it ... nor by the arcs completed in the eccenter}\\
 Namely in this inquiry it is easier to say what is not than what is. The
planet, whenever it is promoted by the sun to the lines from A to C,\ D,\ E, 
\ F,\ G,\ H,
is supposed to assume the distances $ \gamma \alpha,\iota \alpha,\lambda
\alpha, \zeta \alpha,\lambda \alpha,\iota \alpha$, respectively. Now, if its
orbit is a perfect circle, then to equal pieces CD,\ DE,\ EF of the eccenter
correspond unequal descents of the planet on the diameter, namely
$\gamma \iota, \iota \lambda. \lambda \zeta$, and furthermore in disturbed
order, so that not the uppermost are smallest, and the lowest are largest, but
the central ones $ \iota \lambda $ are largest, and the extremes $ \gamma
\iota , \lambda \zeta $ are smaller, and the uppermost $\gamma \iota $ a
little smaller than the lowest $\lambda \zeta$. Namely $ \gamma \kappa$
 and $ \mu \zeta $ are equal, and $ \gamma \iota$ is smaller than $\gamma
 \kappa$, but $ \lambda \zeta$ is larger than $ \mu \zeta$.\\
{\it ... nor by the time elapsed, nor by the angle at the sun, i.e. the true
  anomaly}\\
 And for the same reason $ \gamma \iota, \iota \lambda, \lambda \zeta $ are
not proportional to the times spent in the equal arcs CD,\ DE,\ EF nor to the
angles at the sun CAD,\ CAE,\ 
CAF. The time or the amount of time spent in equal
arcs of the eccenter CD,\ DE,\ EF diminishes continuously from the uppermost to
the lowest arc, the angles at the sun increase continuously, the librations
$\gamma \iota$ however increase towards the central one, $ \iota \lambda$.\par
 Therefore, if the orbit of the planet is a perfect circle, the measure of
the planets descent on the diameter $\gamma \zeta $ is neither the time
nor the space in the eccenter nor the angle at the sun. And these measures
are repudiated by physical speculations as well.\\
{\it ... nor by an imaginated epicycle or eccenter}\\
What, however, if we say the following? Even if the motion of the planet in
an epicycle has problems, the libration could be such that similar distances
from the sun are obtained as those obtained with a real epicycle? \\
 First, we would attribute to the planets own force the knowledge of an
 imaginary epicycle and its effect on obtaining the distances from the sun.
Also would we attribute the knowledge of future velocities caused by the
common motion around the sun. Necessarily the same imaginary change of
velocity in the motion of the imaginary epicycle would be required as with
the real eccenter. This is even more unbelievable as the former concept, in
which the motion of the planet was connected to the knowledge of the 
epicycle or eccenter. Therefore the counterarguments presented before can
be repeated here; the concepts are almost identical.\\
{\it below in chapter 57 the measure of this libration will be disclosed. }\\
 Yet in lack of a better concept, we must presently stop here. The more
absurdities are involved, the easier any physicist will admit below in chapter
52, what the observations will confirm, that the orbit of a planet is not
circular.\\[2.0ex]

\noindent
    {\bf 5.2 By which means or which measure does the planet learn its
distance from the sun?}\\[1.0ex]
So far we dealt with the measure of this kind of libration. What remains
to be done, is to inquire also about the measurement of this measure, i.e.
its size or the motion in space. Namely it is not sufficient for the planet
to know how far it should be from the sun, we should also require that
the planet knows what to do in order to obtain the right distance.\\
{\it Do we have to attribute to the planets quasi a sense for the size of the
  sun? }\\
Whoever is inclined, by this supposition of a perfectly circular orbit,
to attribute an intelligence to the planet, regulating these librations,
cannot say else than this intelligence must be observing the increasing
and decreasing diameter of the sun, and by this means find out at a given
 time the planets distance from the sun. Likewise the sailors cannot find
out from the ocean itself which distance they traversed in the waves since
their path is not marked with any borders. But by the time of navigation, if 
wind and waves have remained constant and the ship never stopped, or by the
direction of the wind and the different altitudes of the pole, or by a
combination of all these informations, or, if the gods like, by the motion
of a system of wheels which are dropped by means of fins into the waves, an
instrument advocated by silly mechanicians who transfer the quietness of
the continents to the floods of the ocean. In the same way, a planet cannot
measure its location or space traversed towards the sun by itself, since only
the pure ether is in between without any signs, but it uses the time or its
equivalent in the same condition of forces, a possibility that was already
denied above, or a mechanical machinery, which is ridiculous ( we assume
planets to be round like the sun and the moon, also it is likely that the
whole ether field moves together with the planet ), or finally some suitable
signs which vary with the distance of the planet from the sun, of which
however none remains except the variable apparent diameter of the sun.\\
 {\it In this way the planets would become geometers which measure their
distance to the sun from a single station, i.e. by the apparent diameter
of the sun.}\\
So we humans know that our distance to the sun is 229 times its diameter,
when the diameter is 30', and 222 when it is 31'. And also, if it were certain
that this movement on the diameter of the epicycle could not be effected by
some material or magnetic force nor by a pure animated force, but would be
governed by an intelligence in the planet, nothing absurd would be stated.
{\it There is something like an intelligence in the planets which respects
the suns body.}\\
That the sun is observed by the planets also in other respects, is apparent
from the latitudes. Namely, since the planets are seen to deviate to the sides
from the central and royal road of this force from the sun like in a torrent,
as stated in chapter 38, they would describe minor circles as seen from the
earth or from the center of the world, parallel to some maximal circle, if
they did not respect the suns position intermediately, and would 
approach or recess in a straight line through the center of the sun. But all
planets describe maximal circles, which intersect the ecliptic in locations
opposite to the sun, as demonstrated for Mars in chapters 12, 13 and 14 above
( see the margin of chapter 63). Therefore also the diameter $ \gamma \zeta $
of the libration is oriented towards the sun and the latitudes respect the sun
everywhere. Below in part V I will transfer also this behaviour of the
latitudes from the action of an intelligence to the action of nature and of
magnetic properties.\\
{\it Possible objections to a sense for the suns body.}\\
{\it 1.The small size.}\\
But don't tell me the diameter of the sun and its variation are so extremely
small that they cannot serve as a ruler. Namely for none of the planets the
sun diameter vanishes completely. At the earth it is thirty minutes, at Mars
twenty, at Jupiter seven and at Saturn three, at Venus however it is forty,
at Mercury planely eighty and up to hundred and twenty. Not about the
smallness of this body one should complain, but about the inadequacy of the
human senses for the observation of such small quantities.\\
Look, how this whatsoever small body is able to move so remote bodies in a
circle, as I demonstrated for the upper planets. Everybody knows about the
illumination of the whole world by such a tiny body. It is then also
believable that, whatever faculty the movers of  a planet have to observe
the suns diameter, that this faculty is so much sharper than our eyes as
their task and eternal motion is more constant than our turbulent and
confuse affairs.\\
{\it 2.The lack of sensual instruments.}\\
Do you then attribute two eyes to each of the planets, Kepler? Not at all.
Nor is it necessary. Nor does one have to give them legs or wings to make them
move. Solid orbits were already excluded by Brahe. And our speculation does
not exhaust all treasures of nature, so that it could be scientifically stated
how many senses exist.Yet admirable examples are at hand. Tell me in terms of
physics by which eyes the animated faculties of sublunar bodies observe the
position of the stars in the zodiac that in case of a harmonic disposition
(what we call an aspect) they jump up and enflame their work? Was it with her
eyes that my mother sensed the position of the stars, to know she was born in
a configuration of Saturn, Jupiter, Mars , Venus and Mercury in sextiles and
trigons, and would therefore give birth to her children, especially to me the
first born, at those days where as many as possible of these aspects,
especially of Saturn and Jupiter, would reoccur or appear in quadratures,
oppositions and conjunctions? What I found true in all examples which occured
to me up to the present day. But what do I deal with these or other equally
absurd matters if not for those who have exercised themselves in nature more
carefully than it is customary nowadays?\\
 Consequently, the one whom we assume here to say that a planets orbit is a
perfect circle, will say with equal right, that the planet effects its
libration by requiring that the sun diameter appears approximately \footnote{
Namely in chapter 57 the proportion will be slightly different}
 in the same inverse proportion to the maximum distance $\gamma \alpha $
as the lines $\gamma \alpha, \epsilon \alpha, \zeta \alpha$
or the equivalent $\iota \alpha, \lambda \alpha, \zeta \alpha$, which
corespond to equal arcs in the eccenter, and by inspection of the sun diameter
the correct distances in the neighbourhood of $ \iota, \lambda , \zeta $ 
at  predetermined times will be obtained.\\
 It should be known, however, that the increase of the sun diameter and the
arc in the epicycle do not well correspond to each other. Therefore the moving
intelligence must have a very good memory in accomodating equal increases of
the sun diameter to inequal {\it sinus versi} of the epicycle arcs; more on
this below in chapters 56,\ 57.\\[1.0ex]
{\bf 5.3. By which means a planet reaches the desired distance from the sun.}
\\[1.0ex]
This shall be enough about the sign of the covered space. It remains that, as a
third item, I spend three words on the physical faculty to transport
 the planet. Whoever says the planet is transported by an inherent force, says
in no way something probable. This we declined in the beginning. But also to
the sun this force cannot be attributed in a simple way. What attracts the
planet will also have to repel it. This contradicts the simplicity of the
solar body. But whoever relates this transportation by some peculiar reasoning
to a mutual agreement between the sun and the planet, will alter the whole
material of this chapter. In this respect a special chapter, 57, will be
devoted to the matter.\\
 You can see, considerate and ingenious reader, that this opinion of a perfect
eccentric circle as a planets orbit involves many incredible things in the
physical speculations. Not because it gives the sun diameter as a sign to the
planets mind --- perhaps also the most correct theory wil do so --- but
because it attributes incredible things both to the mind and to the moving
agent.\\
 But we will have to learn, being really close, how to put these not yet
 perfect, but for the mo\-tion of the sun suitable speculations into numbers.
 It will be found useful for the more accurate invention of the truth, which
is reserved for chapter 57, to have exercised ourselves beforehand.\\[3.0ex]

Here ends chapter 39 of {\it Astronomia Nova}. Before continuing, a few 
remarks may be added which can
help the reader in understanding the text. Kepler likes to use a coordinate
system in which the line joining the sun and the center of the epicycle has 
always the same direction, upwards in the drawings. The purpose of the epicycle
is to provide the  distances between sun and planet, for a given time. It
turns out in the course of the investigation, contained in chapters 42 - 55,
that a more transparent and more correct calculation of the distances is
obtained if one forgets about the epicycle and considers only the line joining
the sun and the planet. When Kepler talks of a libration ( from the Latin
{\it libra} = balance) or balancing movement or upward and downward motion of
the planet on a diameter of the epicycle which points to the sun, he means a
motion on this line. That the line changes its direction in space is not
relevant here. Newton later gave to this line the name of radiusvector (`` the
radius which \mbox{ carries''( the} planet )). Kepler's libration is a
 shortening
or lengthening of this radiusvector, a kind of balancing movement around an
equilibrium orbit which is a circle centered on the sun.\\[2.0ex]

\noindent
   {\bf 6. Astronomia Nova, chapter 56.}\\[3.0ex]
{\bf Demonstration by the previous observations that the distances between
  Mars and sun should be taken quasi from the diameter of the 
epicycle.}\\[1.0ex]
In chapter 46 the width of the moonlet to be cut off the semicircle was found
to be 858 parts if the semidiameter of the circle has 100\ 000 parts, according
to the theory of chapter 45. As I had seen rather clearly from two arguments
presented in chapters 49, 50 and 55, that the width of this moonlet should be
taken only half as much, i.e. 419 or better 432, and around 600 in the scale
in which the semidiameter of Mars' orbit is 152\ 350, I started to think, why
and how a moonlet of this size could be cut off.\par
 While I was anxiously deliberating , being aware that in chapter 45 planely
nothing had been said and my triumph over Mars was futile, I came, perhaps by
chance, across the {\it secans} of 5$^0$ 18 ' which is the maximal optical
equation. When I realized that it is 100\ 429 , I woke up like from a dream
and saw new light. I started to argue as follows. At mean longitudes the
moonlet or the shortening of distances is maximal and has the size of the
excess of the {\it secans} of the 
maximal optical equation over the radius 100\ 000. Ergo, if
at mean longitudes instead of the {\it secans} the radius would be used, the
effect would be what the observations suggest. And, referring to the drawing
of chapter 40 ( Fig.7),
\begin{figure}[hhh]
\hspace*{2.2cm}
  \includegraphics[scale=0.23]{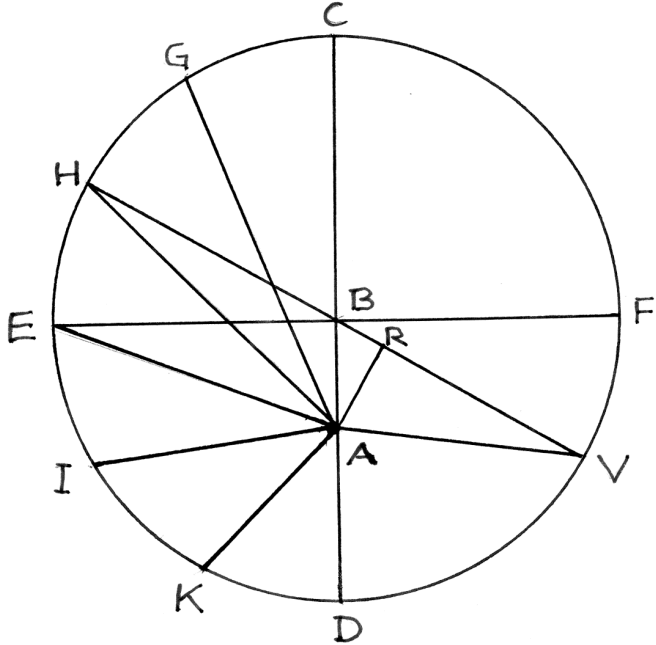}\\
 \caption{Illustration of the idea that instead of the distances 
from the sun EA,\ HA,\ VA for the points  E,\ H, \ V on a circular orbit
the ``diametral `` distances EB,\ HR,\ VR should be used 
( from {\it Astronomia Nova}).}\label{fig:rad}
\end{figure} \\[1.0ex]
 I concluded in general, if instead of HA one would
take HR, instead of VA really VR, and for EA now EB, and so everywhere, the
same thing would happen at all points of the eccenter as what happens
here at mean longitudes. And equivalently, in Fig. 6 instead of the line 
$\alpha \delta $ or $\alpha \iota $ one would have to take $\alpha \kappa $
and instead of $\alpha \epsilon $ or $\alpha \lambda $ one 
would use $\alpha \mu $.\par
The reader is asked to recapitulate chapter 39. There he will find that it
was argued alredy previously, what now the observations moreover confirm,
that apparently the planet seems to perform a kind of libration quasi on
the diameter of the epicycle which always points to the sun. He will also
find that nothing was more contradicting this opinion than that at the time
we were forced by the assumption of a perfectly circular orbit to make the
uppermost librations $ \gamma \iota $ different from the lowest 
$ \lambda \zeta $, while they correspond to equal arcs in the eccenter, namely
the former had to be shorter and the latter longer. Now, if the orbit is 
not circular and  $ \kappa \alpha $ , $ \mu \alpha $ 
are used instead of  $ \delta \alpha $ , $ \epsilon \alpha $ ,or   
$ \iota \alpha $ , $ \lambda \alpha $, as was said, then it follows moreover
that these librations are equal. So, what plagued us in chapter 39 for a long
time, now turns into an argument that we found the truth\footnote{
 The reader will notice that Kepler's conviction to have found the truth
is here linked to the appearance of an unexpected symmetry ( note by the
editor).}. \par
 Also about the fact that the central parts  $ \kappa \mu $ are now larger
 than the outer parts  $ \gamma \kappa $ ,  $ \mu \zeta $, it will be said in
 chapter 57, that this is in agreement with nature, in contrast to what we
 could understand in chapter 39.\par
\vspace*{0.3cm}
The remainder of chapter 56 is skipped here. It deals with a confirmation of
the present hypothesis by observations at all possible longitudes.\par
\clearpage 

\noindent
    {\bf 7. Astronomia Nova, chapter 57.}\\[3.0ex] 
{\bf By which principle of nature a planet is caused quasi to librate
 on a diameter of the epicycle.}\\[1.0ex]
 It appears therefore from most certain observations that a planets orbit
in the ether is not a circle but an oval figure, and that it librates along
the diameter of a small circle, as follows:\\
{\it Definitions. What the cicumferential distance and what the diametral
  distance shall be.
}\\
If a planet after equal arcs in the eccenter assumes the diametral distances 
$ \gamma \alpha $, $ \kappa \alpha $, $ \mu \alpha $, $ \zeta \alpha $ instead
of the circumferential distances $ \gamma \alpha $, $ \delta \alpha $,
$ \epsilon \alpha $, $ \zeta \alpha $ , i.e. $ \gamma \alpha $, 
$ \iota \alpha $, $ \lambda \alpha $, $ \zeta \alpha $ , then it is obvious
that from the perfect semicircle a moonlet is cut off, the size of which at
any location is the difference between the two quantities, say 
$ \iota \kappa $, $ \lambda \mu $ . With this in mind, not due to
 {\it a priori} reasons, but due to observations, as I said, the physical
speculations will proceed more correctly than so far \footnote{The principle
  of this libration is proven to be natural.}. Namely, this libration
accomodates itself to the space covered in the eccenter, but not in a
proportion of rational numbers, that the planets mind would associate equal
parts $ \gamma \kappa $, $ \kappa \mu $, $ \mu \zeta $  of the libration
to equal arcs CD,\ DE,\ EF of the imperfect eccenter --- the parts of the
libration are in fact unequal --- , but in a natural way, which is not based
on the equality of the angles DBC,\ EBD,\ FBE, but on the strength\footnote{
which shall be the genuine and justifying measure of the libration, or the
reason why the {\it sinus versus} of the eccentric anomaly measures this 
libration.} of the constantly increasing angle DBC,EBC,FBC, which strength
follows nearly the so called sine of the mathematicians.Then the ascent is
gradually changed by continuous redution into a descent which is more
probable than a sudden change of the planets direction, what was said in
chapter 39 to be in clear contradiction to the observations. Since moreover
the size of the libration point to a natural process, its cause will also be
natural, namely not  a mind in the planet but a natural or perhaps corporal
faculty. Now, since in chapter 39 we made for very good reasons the assumption
that a planet cannot proceed from one place to another purely by an inherent
force, we have to see if we can attribute the libration in part to a solar
force.\\
{\it Natural examples for this libration.}\\
{\it The oars.}\\
This consideration brings us back to the oars introduced in chapter 39. Let
there be a circular flow CDE,\ IGH ( Fig.8)
\begin{figure}[hhh]
\hspace*{2.2cm}
  \includegraphics[scale=0.27]{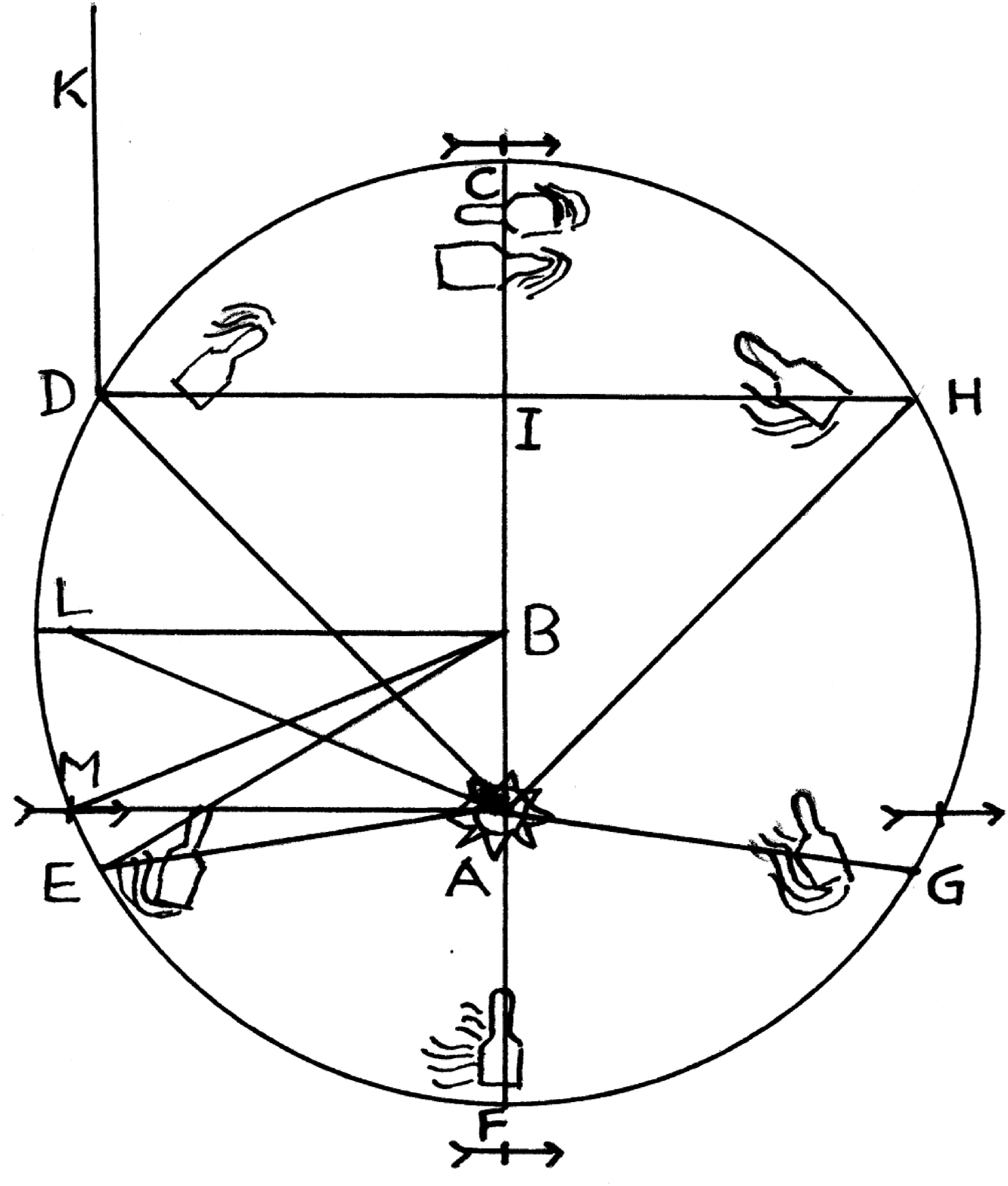}\\
 \caption{
Illustration of two mechanisms which Kepler discusses in order
to explain planetary orbits. The first mechanism assumes a circular flow
around the sun into which the planet dips an oar to bring itself closer to
or further from the sun. The second mechanism is based on a magnetic force.
The planet is represented by a magnetic needle with constant direction,
attracted or repelled by the sun.( from {\it Astronomia
  Nova}).}
\label{fig:oar}
\end{figure} \\[1.0ex]
and in it a navigator who turns his oar once
in two revolutions of the planet by his own very constant force, such that in
C the direction of the oar is perpendicular to the direction of the sun, and
the oar is pointing in turn to the bow or the stern of he ship. In F
the direction of the oar is towards the sun and in the other locations it is
intermediate. The stream in DE will hit the oar such that the ship is directed
toards A; in C the action is only very small since the oar is hardly inclined.
The same is true in F, because the stream hits the oar directly. But in DE the
action is strong because the oar by its inclination is well disposed to an 
approach to the sun. The opposite happens in the ascending semicircle. The 
stream comes in GH from the other side and directs the ship away from the sun.
At the same time, {\it ceteris paribus}, in C the impact will be slower than in
F, because our stream is week in C and strong in F. Also this serves us well
because our libration followed equal spacings in the eccenter and the planet
remained in the upper parts for longer time  than in the lower parts.\\
{\it The defects of this example.}\\
This example teaches only the possibility of the matter. By itself it is less
likely, because the restitution of the oar takes not the same time but twice
as much as that of the stream, and because the faces of the planets as seen
from the earth should be observed to change. The face of the moon however,
which participates with the planets in the motion discussed here, does not
change in its monthly evolution, but is always directed to the earth, from
where its eccentricity is calculated. Add to this that the force of the
stream is material ( the water acts by its weight and material impact ), but
the force of the sun is immaterial. So, other comparisons for the planets have
to be found. They are also not equipped with an oar, a physical instrument,
to take the force of the weights ( which the moving species of the sun does
not have).\\
{\it The example of the magnet.}\\
But from this refutation another perhaps better suited example arises. As the
stream so the oar. The stream is the immaterial species of a magnetic force in
the sun. Maybe the oar has something in common with a magnet. What, if the
planets are somehow gigantic round \mbox{magnets?}
 For the earth ( one of the
planets according Copernicus) there is no doubt. This was proven by William
Gilbert.\\
{\it The magnetic theory of Gilbert.}\\
But let us briefly describe this force. The globe of a planet has two poles,
one which follows the sun, the other which flees it. Let the axis defined by
them be represented by a magnetic needle, the tip of which is attracted by the
sun.\\
{\it It seems that a magnetic disposition in a planet is the reason for
this libration.}\\
It will, however, be retained against its magnetic nature by the translation
of the planet which keeps its axis always parallel to itself, except that in 
the course of centuries the axis points to different stars and causes thereby
a progression of the aphelion. For both of them, in my opinion, a mind may be
necessary, which is sufficiently well instructed for this motion by an
animated faculty, a motion, not of the entire body from place to place ( this
motion was above in chapter 39 correctly taken away from a force inherent in
the planet), but of its parts around a quasi quiescent center.\\
{\it The example of the earth.}\\
Let us examine in the earth an example of such an axis, following Copernicus.
Namely, since the axis of the earth during the annual circuit around a center
remains nearly equidistant to itself for all locations, summer and winter are
caused \footnote{The precession of the equinoxes is similar to the progression
of the aphelia.}. To the extent that very long centuries incline this axis, the
stars are believed to proceed , and the equinoxes to recede.\\
What do we then hesitate, in order to save our ideas about eccentricities,
to attribute to all the planets what we have seen in one of them 
(i.e.the earth) to be the case, according to our understanding of the
precession of the equinoxes and of the suns rise and fall during its annual
revolution ?\\
Here Copernicus was mistaken, thinking a special principle\footnote{For
 Copernicus the natural revolution of the planets around the sun was bound
 like on a wheel or like the motion of the moon around the earth;
 a third motion was therefore necessary to keep the earth's
axis parallel in space. ( note by the editor).} would be necessary for the
earth to librate annually from north to south, such that summer and winter
result, and by the near equality of the libration and revolution times the
small difference between siderial and tropical year would result. All this
results in fact only from the constant direction of the earth's axis and does
not need external reasons, except for the very slow precession of the
equinoxes. So, also here no special advice is neeeded for the promotors of the
planet to transport its body around the sun in a parallel site and 
simultanously perform the libration. One depends on the other in a natural
way. Only the progression of the aphelia remains to be reflected upon.\\
{\it The reason why  the libration is fastest in the middle.}\\
Now, when the needle is  in C and in F, there is no need for the planet to
approach or recess, since the ends of the needle have the same distance from
the sun, and the planet would direct the tip of the needle towards the sun,
if not retained by that force which keeps its axis parallel. When the planet
leaves the point C, gradually the front will approach the sun, the tail will
tend away from it. Gradually the planet starts to navigate towards the sun.
Behind F gradually the tail approaches and the head leaves the sun. Gradually
then the whole body , by natural hatred, flees from the sun. Near A, where
the axis points directly to the sun, here the approach, there the flight is
strongest. This is really what our assumptions, as deduced from observations,
did require, where out of the parts of libration 
 $ \gamma \kappa $ , $ \kappa \mu $ , $ \mu  \zeta $ , corresponding to 
equal arcs in the eccenter, the middle parts , $ \kappa \mu $  were very
long,  but the parts towards  $ \gamma , \zeta $ tiny.\\        
{\it The reason why the libration at top is slow, at bottom fast.}\\
But also this agrees, that the observations want $ \gamma \kappa $   and
$ \mu \zeta $ to be equal, while their arcs $ \gamma \delta $ , $ \epsilon
 \zeta $ or rather CD,\ EF in the eccenter are equal but are completed in
 unequal times, CD taking longer, such that the piece $ \gamma \kappa $  of
the libration is completed more slowly than the equal piece $ \mu \zeta $ .
Namely also the magnets approach each other more slowly from a large distance
and faster from a small distance.\\
{\it The axis of force is retained parallel in a planet by a natural force
.}\\
Also the force which keeps the magnetic axis parallel, disregarding the
direction to the sun, can now be transferred from the occupation of a mind,
as we thought a moment ago, to the work of nature.\\
{\it With an exception, however.}\\
 Although there seems to  be the
obstacle that nature acts always the same way, while this retaining force is
seen to be different at different times, like the inclination of the axis to
the sun, to which this force has to be compared, is nearly vanishing at mean
longitudes but is very strong at the aphelion and the perihelion.
But what impedes this retaining force to be many times stronger than the
inclination of the axis to the sun and to become hardly or not at all
fatigated by such a weak adversary?\\
{\it The  magnet as an example.}\\
Again let us take the magnet as an example. In a magnet manifestly two forces
are mixed, one which directs to the pole, the other which aims at the iron.
So, if the compass needle is directed to the pole, the iron is reached from
the side; the compass deviates a little from the pole and is inclined towards
the iron, due to some familiarity with the iron, but points predominantly to
the pole.\\
{\it The  reason why the magnet deviates somewhat from the pole.}\\
This happens according to Gilbert because the compass is deviated by
continents of particularly large size; and so the reason for this declination
is in the configuration of the land masses, of which, to the right or to the
left, higher, larger and more powerful ones may be in the neighbourhood.\par
  In this respect we can allow to both natural faculties to act on the same
subject in an equalized manner and thus show the rather clear and by no means
obscure reason for the translation of the aphelia in the moderated action
of the two forces.\\
{\it The reason for the motion of the aphelia.}\\
 Let it be that this force to direct the axis towards the sun yields a little
to the retaining force, in proportion of their strengths. In the semicircle of
the aphelion like in C, the tip of the needle is inclined a little towards H,
that means clockwise, the tail points away from the sun, winning somewhat
over the retaining force. Therefore the aphelion will become retrograde. But
in the semicircle of the perihelion, like in F, the same tip will be inclined
towards G, i.e. anticlockwise, again winning over the retaining force in the
contrary direction. So the aphelion will become rectograde and fast. Because
AF is shorter than AC, and the sun is closer to F than to C, also the force
directing the magnetic axis towards he sun will be stronger in F than in C.\\
{\it Why the aphelia do not recess.}\\
The anticlockwise inclination at perihelion will therefore not only compensate
the clockwise inclination at aphelion but will surpass it.
So this is the reason why the apsides progress and do not recess. The aphelion
which we found will therefore be equally valid in the true anomalies of 90
$^0$ and 270 $^0$, where the axis of force is directed towards the sun
, which is
its natural direction. The motion of the aphelia will be in the form of a
spiral, like the motion of the precession of the equinoxes, but for a
different reason, as will be shown below in chapter 68. The parallel direction
of the 
magnetic axis or the corresponding force, its guard, will not respect these or
those stars, but only the location of the planets body, at any time. And, if
you think it over, simply  because this direction is more similar to quietness
than to motion , in matter, and is with more right attributed to some
disposition of the body than to some mind.\\

{\it It shall be agreed that a correctly positioned magnet will perform these
 librations.}\\
Let us then follow this similarity of a planets libration with the motion of
a magnet more closely and this with a very fine geometric proof,
to make it clear that magnets have the same motion as we learned for planets.
Let DFA (Fig.9)
\begin{figure}[hhh]
\hspace*{2.2cm}
  \includegraphics[scale=0.23]{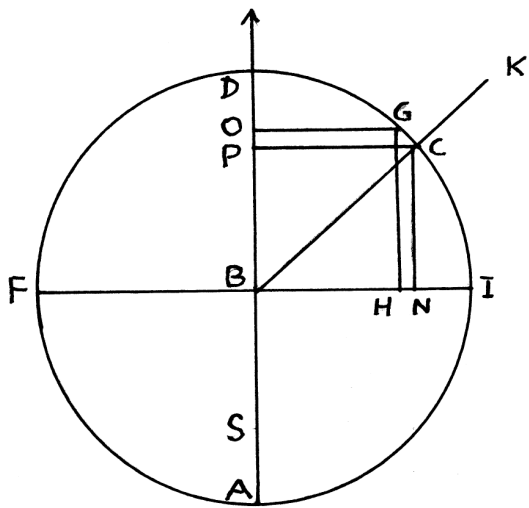}\\
 \caption{
 Illustration of how a planet with magnetic fibers in the
direction AD is attracted or repelled by the sun located in the direction K
( from {\it Astronomia Nova}).} 
\label{fig:magnet}
\end{figure} \\[1.0ex]
be a round magnet or the body itself of Mars, DA the line
in which the magnetic force acts, D the pole which aims at the sun, A the pole
 which flees it. First note, that in this speculation it amounts to the same
if we consider the entire magnetic body or only one physical line of force,
 parallel to DA.\clearpage 
\begin {center}
{\it Coeca mari signas Nautis vestigia Magnes\\
 Quid mirum Errones nutibus ire tuis ? \\[1.0ex]
 Hidden signs you offer, magnet, to the sailors\\
What wonder that erratics follow now your traces ?}\\[1.0ex]
\end {center}
Since now this magnetic force is corporal, and dividable with the body, as
proven by the Englishman Gilbert, B. Porta and others, certainly, because
a globe consists of an infinite number of physical lines parallel to DA, whose
force  tends in one and the same world direction, about the individual
lines separately the same judgement as to their way of motion will be made as
for all of them joined together, and {\it vice versa}. Let therefore instead
of the whole body and all its filaments the middle axis DA be proposed for
argumentation. Let DA be bisected in B, and FBI drawn perpendicular to DA.
If the planet is located such that BI points to the center of the sun, there
will be no approach. The angles DBI and ABI are equal, and therefore of the 
same strength, this for approach, that for recession. This is the same as
equilibrium in mechanics. So B, the center of Mars, will remain in one of the
apsides, say the aphelion, farthest from the sun. Let us take some arc IC as
a measure of the true anomaly and draw BC which points towards K.
 Let the planet
be located in such a way that BC points towards the sun which is understood by
K. First the measure of the planets strength of access shall be inquired.
There will be approach, because the pole D which aims at the sun, is inclined
towards the sun in K by the angle DBK. The pole A however flees the sun by
the angle ABK. Because the strength of the angle is natural, it will be in the
ratio of the balance. The ratio between DP and PA , where CP is the
perpendicular to DA through C, will be the ratio of the balance. A balance 
suspended along the line BK with the arms remaining at the angle DBK will
have the weights of the arm BD and the arm BA in the ratio DP to PA; in the
same way as if he arms would be suspended in P and the weight of BA would
correspond to PD, and the weight of BD to PA, then the line DA would be
perpendicular to the balance line CP. See my ``Optics'' and do not trust
experiments which are not done carefully. So, as DP to DA is the ratio of
the strength of the angle ABC to that of the angle DBC. The force of recession
is therefore measured by DP, that of access by PA. Subtract from PA the amount
DP which gives AS. Ergo, SP is the measure of the  force of access to the sun,
 corrected for the effect of recession, and this in proportion to AD as the
 maximum force. But as half of this, DB, measures the maximum force, also PB,
half of PS, that means the sine CN of the true anomaly CBI, measures the pure
force of access at this position of planet and sun. So the sine of the true
anomaly is the measure of the strength of access
 of the planet to the sun at this
location. And this is the measure of the increment of the force.\\
{\it What is the measure of the space covered by the libration up to a given
moment?}\\
The measure of the space covered by these continuous increments of the force
is very much different. The obserations show that IH, the {\it sinus versus}
of the arc GI, is the measure of the total oscillation, if GI is the 
eccentric anomaly corresponding to the true anomaly IC. This could have been
also deduced from the previously indicated measure of the velocity CN, but
now we brought the experience into agreement with the idea of the balance. 
Since the sine is the measure of the strength of every angle, the sum of sines
will nearly equal the sum of the strengths or impressions of all equal parts
of the circle, the common effect of which is the total achieved libration.
And the sum of sines IG ( let the otherwise different anomalies IC and IG be
equal to avoid confusion ) is approximately in the same proportion to the
sum of sines of the quadrant as IH, the {\it sinus versus} of this arc IG to
IB, the {\it sinus versus} of the quadrant. I said approximately. Namely in
the beginning, where the {\it sinus versus} is small and has small increments,
it is only half the size of the sum of sines. Look here. Let the quadrant of
90 $^0$ have 90 parts; the sum of 90 sines is 5\ 789\ 431. Already long ago
I added them in order.\\ 
{\it What is the proportion of the sinus versus of any arc to the sum of sines
 of all previous degrees? }\\
The sum of the sines in the arc 1 $^0$ , i.e. the first sine, is 1745, and in
the proportion to the previous sum, as 100\ 000 to 30. In contrast to this,
the {\it sinus versus} of 1 $^0$ is 15, which is half of 30.\\
{\it The ratio is practically constant, with negligeable error.}\\
The reader should not be deterred by this unmathematical and wrong
start. Before the amount of libration becomes significant, the two calculaions
have negligeable difference. The sum of 15 sines which is 208\ 166 indicates
3594. But the {\it sinus versus} of 15 $^0$ is 3407/100\ 000, a little less
than the previous number. So the sum of 30 sines, which is 792\ 598 indicates
by the law of proportions a libration of 13\ 691 in 100\ 000, but the 
{\it sinus versus} of 30 $^0$ gives 13\ 397. And the sum of 60 sines, which
is 2\ 908\ 017 indicates a little more than 50\ 000, while  the {\it sinus
  versus} of 60 $^0$ is 50\ 000.\\
{\it The application of the demonstrated magnetic libration to the observed
libration of a planet.}\\
Since it is now demonstrated, if a magnet is positioned  as we assume the
planets to be positioned  with respect to the sun, that its libration is
measured by the {\it sinus versus} in what concerns the covered space. The 
planets are observed to librate by the same amount of the {\it sinus versus}
of the eccentric anomaly. It is therefore very plausible that the planets are
magnetic and dispositioned to the sun as we said.\\
{\it The relation betwen the various sinus versi of the eccentric anomaly
is the same as the relation between the sum of the sines of the corresponding
true anomalies, very accurately.}\\[1.0ex]
Let us demonstrate now that I made no big mistake in assuming the arcs IC and
IG to be identical. When I say, the arc IC is for the planet the measure of the
true anomaly, I speak correctly, and then CN is the genuine measure of the
strength which acts on the planet when the sun is on the line BK. If I say,
however, IG is the measure of the eccentric anomaly, I speak incorrectly,
because I use the circle of the planet to represent the eccenter. But since
in the descendent semicircle of the eccenter a larger arc of the eccentric
anomaly corresponds to a smaller arc of the true anomaly, i.e. IG corresponds
to IC, we collect more sines in IG than in IC, and rightly so.\\
{\it The later a planet is in any arc, the smaller the portions of the true
anomaly have to be made, in order that their collected sines become the
correct measure of the force emanated at this true anomaly. }\\ 
Namely because the sine measures the strength, and the strength acts according
to time and according to the distance to the sun ( the closer the magnets the
stronger they are), that means, to be brief, according to the arc IG, as many
sines have to be included in IC as are found in IG.\\
Only in that respect we made an error, that we took these many sines larger
than they are, as GH is larger than CN. But this excess is in the first place
by itself extremely small  and unnoticeable. Namely at the beginning of the
quadrant the arcs IC and IG hardly differ, and the sines are small, at  the
end of the quadrant, where the equation of the eccenter CG is at its maximum, 
the sines differ very little.\\
Finally this error is to our advantage. The sum of the sines is always a little
larger than the {\it sinus versus}, to which we try to accomodate and
reconciliate the amounts of libratory and magnetic deflections. Ergo, this
our present error to accumulate larger instead of smaller sines, is covered
if we use the {\it sinus versus} instead of the sum of the sines. The sum
of the sines is not exactly the same as the {\it sinus versus}, but exceeds
it, as far as the librations are concerned.\\
{\it The defect in the proportion of the  sinus versus and the sum of
the sines is compensated by the contrary error that we collected the
larger sines of the eccentric anomaly instead of those of the true 
anomaly. }\\
So we brought the matter with very good reasoning to an end within the limits
of sensitivity. In conclusion, a planet, like a magnet, accesses and flees the
sun by the law of balance, along an imaginary diameter of the epicycle which 
tends towards the sun and the real and force diameter DA of the planet points
to mean longitudes, namely BD at our time to 29$^0$ {\it tauri}, BA
 to 29 $^0$
{\it scorpii}, since the aphelion is in 29 $^0$ {\it leonis}.\\ 
{\it The magnetic force in the planets is excited and activated by a similar
force in the sun. }\\  
In this way the access by libration is achieved not by the action of a mind,
but due to an inherent and solitary magnetic
force whose definition however depends on the foreign body of the sun. The
force by definition aims at the sun or flees from the sun. Although this
connecting force  between magnets has to be mutual, I denied above in chapter 
39 the force of the sun to the planets to be attractive, or only to the extent
that is required by the argument used. Here, however, the force is assumed 
to be simultanouly attractive and, at another site, repulsive. Also this is
assumed, that the sun, like the virgin iron, is only aimed at but does not
 aim itself, since its filaments were above supposed to be circular, those 
of the planets however, are here taken to be straight.\\
{\it  The difficulty and imperfection of this example of the magnet }
\\[1.0ex]  
It is sufficient for me to have demonstrated the possibility of the matter
in principle, by this example of a magnet. Incidentally, about the matter
itself I am in doubt. For what concerns the earth, it is sure that its axis, by
the constant direction of which the seasons are created, is not suited for
this libration and for the definition of the aphelion. Because the apogee of
the sun or the aphelion of the earth nowadays nearly coincides with the 
solstitials, and not with the equinoxes, what would suit us. Nor does this
axis remain in the same distance to the cardinal points. But if this axis is
unsuited, none in the whole body of the earth is suited, because there is no
line which remains at the same place, when the whole body turns around the
previous  axis in its dayly untiring revolution.\\
{\it About the mental principle of this libration. I hesitate to 
call it rational to avoid confusion with a rational discourse. }\\  
But in fact, if no planely material or magnetic faculty can achieve the task
given to each planet separately, due to the lack of the proper means, that
is to say a diameter in the body which remains to itself always equidistant
during translations, which is a defect that already appears in one of them,
namely the earth, then we have to call for a mind, which, as was said in
chapter 39, gets to know the distances covered by looking at the increasing
diameter of the sun; this mind is to direct the faculty of the planet, be
it natural or animated, to accomodate the planet in a parallel site, such that
it is driven by the force of the sun in the required way and librates with
respect to the sun ( Namely a mind alone without a faculty of inferior grade
 can by itself do nothing in a body). At the same time the mind is adviced to
make the times of periodic restitution and libration not just equal and so to
transfer the apsides. The likelihood of these things was explained in
chapter 39.\par
 It remains, since we know already from observations the laws and size of this
 libration, by which the apparent diameter of the sun varies, --- laws which
we did not yet know in chapter39 ---  that we find out if these laws are such
that it is likely, that the planet can get to know them. The laws of the
libration are that the {\it sinus versus} of the eccentric anomaly measures
the part of the libration which has been completed.\par

{\it The increases of the sun diameter are proportional to the 
 sinus versus of the true anomaly. }\\   
I state therefore in the beginning: given that, in agreement with observations,
a planet after equal arcs in the eccenter is found in the points
$\gamma, \kappa, \mu, \zeta$  and not in $\gamma, \iota, \lambda, \zeta$,
then the correct measure of the increase of the sun diameter is the  
 {\it sinus versus} 
of the true anomaly \footnote{The {\it sinus versus} of the
 eccentric anomaly measures the libration of a planet.\\The {\it sinus versus}
 of the true anomaly measures the increase of the sun diameter as seen by an
 observer in the planet.}; we also know that the {\it sinus versus} of the
 eccentric anomaly measures the libration.\\
 Since therfore the mind of a planet, if there is one, realizes the space
 covered by the libration only by the increased diameter of the sun, as said 
in chapter 39, it will necessarily have to get knowledge of the  {\it sinus
  versus} of the true anomaly, in order to increase accordingly the sun
diameter.\\
 The proof of this is here. Let the planet after equal arcs CD,\ DE,\ EF of the
imperfect eccenter be in $\gamma,\kappa,\mu,\zeta$ . The point D and H shall
be connected; the diameter CF is intersected in I. Since $ \delta \kappa
 \theta$, $ \epsilon \mu \eta $ are straight lines, they intersect the epicycle
in similar arcs as in the eccenter, by construction. So $\gamma \zeta $ to
$ \gamma \kappa $ will be in the same proportion as CF to CI; one is the 
measure for the other.\par
 Because this is so, I claim, that consequently the increments of the diameter
 of the sun in $\alpha $ as seen from $ \gamma, \kappa, \mu, \zeta $ will be
accumulated by the same amount in which the  {\it sinus versus} of the true
anomaly increases. To prove this in general would be not appropriate. It can
however easily seen to be true in general, if we prove it for the center
and for the extremes.\par
 In C the true anomaly is zero, the  {\it sinus versus} is zero, and the sun,
seen from $\gamma$, appears smallest, such that also its increment is zero. In
F the true anomaly is 180 $^0$ , the  {\it sinus versus} equals the full
diameter 200\ 000. And the sun, seen from $\gamma$, appears largest, so its
increment will have reached the total.\par
 As to the true anomaly of 90 $^0$ : the perpendicular to CF in A intersects
 the eccenter in M; the tangent from $ \alpha $ to the epicycle contacts the
 epicycle in $\nu$ (Fig.10)
\begin{figure}[hhh]
\hspace*{2.2cm}
  \includegraphics[scale=0.27]{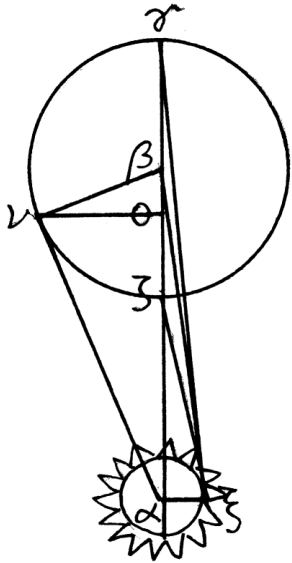}\\
 \caption{
Drawing by which Kepler proofs that the diameter $\alpha \xi$
 of the sun appears midway between extremes, if the distance of the planet
from the sun is o$\alpha$, or else if the planet is located in position
$\nu$ of the epicycle, where $\nu$ is the point in which the tangent from
$\alpha$ touches the epicycle ( from {\it Astronomia Nova}).} 
\label{fig:sin}
\end{figure} .\\[2cm]

Because $ \alpha \nu \beta $ is a right angle according
to Euklid III,22 and MAB is a right angle by construction, and $\beta \nu$
and BA are equal by construction, and so $\beta \alpha$ and BM, the triangles
are equal equal and congruent, therefore also the angles $\nu \beta \alpha$
and ABM are equal. The perpendicular to $\gamma \zeta $ from $\nu$ intersects
 $\gamma \zeta $ in o. Therefore, since $ \nu  o  \beta $ is a 
right angle and also MAB, $\nu\beta $o  is equal to MBA and the triangles
are similar, so $\nu\beta$ to $\beta$o is as MB to BA and 
{\it vice versa}. Because $\nu \beta, \beta \gamma,\beta\zeta$ are equal and
also MB,\ BC,\ BF, the joint of $\nu\beta$ and $\beta $o
 , i.e. $\gamma  $o,  relates to  o$ \zeta $ as the joint of MB and
BA, i.e. CA , to AF. Since CA is the  {\it sinus versus} of the eccentric
anomaly CBM, and is understood to measure the corresponding part of the
 libration, $\gamma$o will be this part. Ergo at the eccentric anomaly
 CBM, or the true anomaly 90 $^0$, the planet will be in o.\par
But the {\it sinus versus} of the true anomaly 90 $^0$, i.e. CAM, is half of 
the total diameter, i.e. 100\ 000. I claim that also the size of the diameter
 of the sun , seen from o, will be half way between the sizes seen from
$\gamma$ and from $ \zeta$, such that falf of the increase is reached, when
the planet is in o below $ \beta$.\par
 Let the diameter of the sun be $\alpha \xi$, and the angles of vision 
$\xi\zeta\alpha$, $\xi$o$\alpha$, $\xi\gamma\alpha$. Now AF and $\zeta\alpha$
are equal, also AC and $\alpha\gamma$. CA to AF is in the same relation
as $\gamma$o to o$\zeta$. Ergo $\gamma\alpha$ to $\alpha\zeta$
is as $\gamma $o to o$\zeta$. But $\gamma\xi$ is unnoticeably different from
$\gamma\alpha$  and so is $\gamma\xi$ from $\xi\alpha$. Ergo $\gamma\xi$ to
$\zeta\xi$ is practically as $\gamma $o to o$\zeta$. In the triangle
$\gamma\xi\zeta$ , the angle at $\xi$ is divided by the line $\xi$o such that
the basis $\gamma \zeta$ is divided in proportion of the sides $\gamma\xi$,
 $\zeta\xi$. Ergo by the inversion of Euklid VI.3, the angle $\gamma\xi\zeta$ 
is divided by the line $\xi$o in two equal parts, and  $\gamma\xi$o is half
 of  $\gamma\xi\zeta$ , the full increment of the sun diameter, {\it quod
erat demonstrandum} . So for the extremes and the center it is demonstrated,
that, if the diameter of the libration is divided by the planet in proportion
of the {\it sinus versi} of the eccentric anomaly, then the diameter of the
 sun will increase in proportion to the {\it sinus versi} of the true
 anomaly.\par
 This is apparent also from the following argument which is included for
 further evidence: Let the perpendicular to CF in B be BL, where L is the
 intersect with a circle around A with radius BC. Since CBL, the eccentric
anomaly, is 90 $^0$, the {\it sinus versus} of it will be 100\ 000, half of
 the full diameter, therefore the libration  $\gamma\beta$ will be half of the
 total  $\gamma\zeta$ , and the distance will be  $\beta\alpha$ . This is
equal to AL, by construction, so the planet will be in L. And since AL equals
 BC or BM, BA is the common side and LBA like MAB are right angles, the
triangles BMA and ALB are congruent. So also BL equals AM. But AM is the same
as  $\alpha\nu$, as above, and so is BL. But  $\alpha\nu$, the hypothenuse of
the triangle  $\alpha$o$\nu$ is larger than $\alpha$o, ergo BL is larger 
than  $\alpha$o, and AL larger than BL, therefore AL is much larger than
 $\alpha$o. The sun therefore appears smaller in the distance AL than in the
distance $\alpha$o. The distance $\alpha$o was already seen to be midway
between the maximum and minimum, therefore in the distance AL the sun
appears smaller than on average.\par
Even if in L half of the semicircle is absolved, less than half of the
 increment of the sun diameter is completed. Also because the true anomaly LAC
is less than half, 90$^0$. And this is what tormented us in chapter 39, as
 said in the preceeding chapter 56. Namely if the planets orbit would be a
 perfect circle, the increase of the sun diameter would measure the increases
of the {\it sinus versi} of the eccentric anomaly, the observation of which
is more foreign to the mind than the observation of the true anomaly, as we
heard already. See her from the contrary how convenient this measure is to the
 planet and how plausible.\par
{\it The planet cannot have knowledge of the eccentric anomaly }\\   
If we would declare the measure of this libration, namely the {\it sinus
 versus} of the eccentric anomaly, as recommended by the observations,
 to be comprehensible to the planet, we would take away from its mind the means
of the variable sun diameter, because this diameter does not accomodate itself
to the {\it sinus versi} of the eccentric anomaly. The planets orbit is namely
 not a circle. And the mind of the planet would have to find out the parts of
 libration, or the spaces to be covered, without sign, what we already called
 absurd; he would also have to find the eccentric anomaly, that is the angle
 between two straight lines from the center of the eccenter, one through the
aphelion, the other through the center of the planet. In Fig.8 it is the angle
DBC ( or the complement of KDB, where K is a straight line from D parallel to
 BC). If the mind perceives the angle KDB, it is necessary that it perceives
the three points K,\ D,\ B. For D there is no doubt, because D is the center of
 its own sphere. About K , I have little doubt. BC and DK practically coincide
due to the infinite distance of fixed stars into the same location among the
 stars, and the stars are real bodies.\\

{\it This reasoning was not necessary in the natural mode discussed a 
little before. }\\   
Therefore there is nothing wrong with the planet looking by some unknown sense
 to that fixed star which acts at the time as a host for the aphelion. Only
for B it has to be denied that it can be sensed by the planet, because it is
not endowed with a body.\par
  Besides, if the reason disappears, why B should be inspected, also the
 effect goes away. But B has to be inspected, if a circle should be 
accomplished. The planetary orbits, however, are not perfect circles, what was
proven in chapter 42 from observations. Ergo the planets do not focus on B. 
And when B is quasi the center, it is later than the path CD. But if it is to
 be inspected by the planet, it has to be earlier.\par
 For these reasons I deny, that the 
{\it sinus versus}
of the eccentric anomaly gives to the planet the measure of its libration, not
 because it is not the measure, but because, although it is the measure, it
cannot be observed by the planet. But if we give to the planet the increasing
 and decreasing diameter of the sun as measure or a support by which it finds
the correct and by themselves imperceptible distances during the libration,
and state the true anomaly , in Fig.8 the angle DAC or rather KDA, to be
the rule and measure  which the planet should observe in order to vary this
 sun diameter, according to the proof given a moment ago, then we are on the
 right track.\par  
{\it The planet can know the true anomaly. }\\   
 Namely both signs are perceptible: as far as the part of the libration is
 concerned, the growing and decreasing diameter of the sun, for the
 measurement of the angle three points endowed with bodies. Namely in A is
the sun itself, In D the planet and in K the stars near the aphelion.\par
 Perhaps we have to say ( what we considered already in chapter 39 above, in
 case the forces of nature be not sufficient to govern the celestial motions)
that the planets have to be given a sense for the light of the stars and the
 sun such that they can estimate this angle of true anomaly from the rays which
concur in the center of their bodies.\par
 {\it If a planet has a sense for the true anomaly, it does not estimate
the angle, but the sine of the angle. }\\
 Only one difficulty has to be removed. Why is not the angle itself a measure 
for the task of the planet which is to increase the sun diameter on approach,
but instead of the angle its {\it sinus versus}\footnote{Shortly before the
 sine of the eccentric anomaly( or the corresponding true anomaly) indicated
the strength of libration, but the {\it sinus versus} of the eccentric anomaly
indicated the achieved libration; here the sine of the true anomaly indicates
the velocity at which the sun diameter increases, the {\it sinus versus} of
the true anomaly however indicates the total achieved increase by all
 preceeding velocities.}? And by which means does the planet perceive the sine
of the true anomaly? Does it proceed like humans by geometric reasoning? So
 far, however, in administrating celestial motions, no task had to be given to
 planet which could not be dealt with by a divine instinct, prevailing since
the very begin of the world up to the present day, beyond any calculational
effort. Let us repeat, as we just said, that the sine of the true anomaly
indicates the strength of the angle KDA, of which Aristoteles talks in his
mechanics, and also we , in this same chapter shortly before. Namely two arms
at an obtuse angle are easier to direct than when at right angles, and this
in proportion to the sines. And conversely, two arms joined under an acute
 angle are easier to force to a straight line, where their ends join, than when
at a right angle. Repeat the proof with the premisses given above.\par
 So in one way there is nothing absurd if we say ( in our human way of
 understanding) given that planets have a sense for the strength of an angle,
that they get  to know the sines of the angles. But why does the sine measure 
the natural strength of an angle ? No wonder we are back to the principles of
 nature. Let there be, as previously, certain fibers in the planet, in which
a magnetic force sits along a line which points to the sun. Let also that be
 attibuted , not to the nature of the body, as before, but to an animated
 faculty or one which reigns over it from inside, that it directs this magnetic
 axis always to the same stars while it is seized by the sun, except that in
 the course of centuries the axis is slightly deviated. So here arises a
 conflict between the animated and the natural faculties, and the animated one
wins. Nothing else was said in chapter 34, when we argued that the planets
 naturally like to rest but are moved by the external force of the sun.\par
 Or take a more suitable example. The natural weight of the human arms points
downwards to the center of the earth. The animated faculty of a banner carrier
, however, enables him to erect the banner over his head and to swing it
 around. Here the animated faculty wins over the  natural weight and it would
 win forever if the man's body with all its faculties would not be mortal.\par
 With these asumptions a planet will be able to understand and perceive the
 strengths of angles due to a struggle between the animated faculty which is
 prepared to retain the magnetic axis and the magnetic force in direction of
 the sun.\par
 This view seems also to be confirmed by the example of the moon , which is
certain to be accelerated when in line with sun and earth, maybe by this 
strength of angles.\par
 Finally , the conclusion will be this:\\
{\it The type of celestial motions, if supported by a mind. }\\
A planet in the aphelion is not inclined towards the sun but is moved forward
according to the distance AC; following this promotion it gets to the angle
KDA. In proportion to the strength of this angle the planet  itself increases
the diameter of the sun, approaching the sun. By this approach it diminishes
the distance, which becomes AD. Due to the minor distance the planet moves
 faster. Faster also changes the angle KDA, and faster the planet increases
the semidiameter of the sun ( {\it ceteris paribus}). In this way a perennial
circulation is achieved, not by intervals, as we assumed in our reasoning and
 in the calculations, but completely continously.\par
{\it Comparison of the mental and the magnetic principle. }\\
 I said this so far under the condition that, if necesary, we have to resort
 to a mind, if the libration about which we know from observations cannot be
achieved by some magnetic force in the planets. Incidentally, if one wants
to compare this natural and that mental motion, the former stands {\it per se}
and needs nothing, the latter, however it is endowed with an animated faculty
 to move the body, seems to confirm the former and to require its help. Namely
in the first place a mind by itself cannot do anything in a body. It is 
necessary to add a faculty to execute its projects in the librating planet.
This faculty will be either animated or natural and magnetic. Animated it
 cannot be. Namely an animated faculty cannot transport its body from place
to place ( as required in this libration) without the help of another body.
So the faculty will be magnetic, i.e. a natural consense between the bodies
of the sun and the planet. So the mind calls to nature and to magnets for
 help.
Thus the mind, when half the way through with its work which consists in
increasing or decreasing the diameter of the sun according to the size of the
true anomaly, has completed in the upper half the longer part $\gamma$o of the
libration and in the lower half the smaller part o$\zeta$. Neither  $\gamma$o
nor o$\zeta$ correspond to the time spent. For   $\gamma$o more time  is
 needed than what its excess over o$\zeta$ would require. Also the time
 intervals do not continuously increase from $\zeta $ towards $\gamma$; for
$\gamma \kappa$ less time is needed than for $\mu \zeta$ . But the works of 
a mind use to be constant.\par
$\:$

 Therefore we had to endow the mind with an animated and a magnetic faculty
and to install a struggle between them to remind him of his duty, what
neither the equality of the time spent or that of the space covered
 could have done. So again we asked for help from nature.\par
On the other side, all these modifications are in fact inherent to the work of
the external magnetic force of the sun in conjunction with a magnetic force in
the planet, as explained above. If these magnetic forces do all the work by
themselves, what is the direction of a mind needed for ?\par
 Also, if we were uncertain about a magnetic force inside the planets, 
contemplating the axis of the earth which is not on the suns line of apsides,
this difficulty is in fact common to both explanations. Namely, assuming a
 mind, we are nevertheless forced to admit an axis such as we want it for the
 earth, by means of which the mind learns about the strength of the angle, or
its {\it sinus versus}. Against that idea we are vehemently urged by
 probability to ascribe this libration of the planets, which follows beyond
 any doubt the laws of nature, to ascribe it totally to nature, however it
may manifest itself in the planets.\par
And this so much that I do not know if I have made this sensual comprehension
of the sun and the stars which I reluctantly accept myself and attribute it to
a mind in the planet, if I made it sufficiently plausible to 
a philosophically minded reader.\par
 In addition, there seems to be even in the methods which we attributed to a
 mind ---- the most reliable ones of all  --- a certain geometric
 uncertainty of which I do not know if it is not repudiated by God himself,
 who so far was  always found to proceed in a mathematically demonstrable way.
Namely, if a planet in order to approach the sun , partly by an inherent
 force, comes degree by degree of the solar force closer to the sun ( as it
 does) and if these different degrees reciprocally strengthen its own force of
 approach by increasing the angle which is assumed to determine the law of
approach, or else the sun diameter, the degree of  approach  will be given
to some extent
by the approach itself, and will be, in the intention of the planet,
 simultanously beforehand and afterwards. Since the approach is by unequal
 pieces, it will need a measure. The result will then be 
given not by a direct calculation, but quasi by the {\it regula falsi},
since both forces influence each other and develop themselves at the same time
in the same revolution of the bodies.\par
\vspace*{1.0 cm}
Here we may stop with Kepler's text. The remainder of chapter 57 is devoted to
ideas on precession.\par Since the last paragraph really deals with the
question that the integration of differential equations may be complicated, in
words which are written decades before the invention of the calculus, the text
will be repeated here in the original Latin version for the benefit of readers
who will appreciate. It is also a fine example of Kepler's literary style.\par
\clearpage
{\it Accedit et hoc, quod in ipsis etiam modis, quos Menti praescripsimus,
omnium, qui possunt esse, probatissimis, implicari videtur quaedam incertitudo
Geometrica; quae nescio an non a Deo ipso repudietur, qui hactenus semper
demonstrativa via progressus esse deprehenditur. Nam si Planeta prout ad
Solem, partim vi insita approquinquaverit, in alium et alium gradum virtutis
ex Sole adventitiae venit ( ut quidem venit) et si diversi gradus, reciproce
ipsius etiam Planetae vim appropinquandi intendunt, dum angulum augent, qui
Regula ponitur appropinquationis, seu auctionis diametri Solis: Nisus Planetae
proprius, denique sibiipsi fiet ex parte mensura, et in intentione Planetae,
simul prius et posterius; cum sit per partes inaequales, et ob hoc ipsum 
mensura indiguerit. Quo pacto non demonstrative, sed quasi per regulam falsi,
dabitur exploratio temperandarum virium utriusqus virtutis, ut eodem tempore
sese expediant, eodem corporis circumactu.}\\[2.0ex]

\noindent
{\bf 8. A concluding comment. }\\[3.0ex]
 Perhaps it is not totally inappropriate to add here a final remark. It looks
like it might be possible to set up differential equations of motion without
knowledge of the forces. This, however, would be a wrong conclusion. What
Kepler was missing is the connection between force and acceleration.
Acceleration is a quantity not immediately obvious for an observer. Its
relation to force could only be established by quantitive measurements ---
more specifically by the measurements of Galilei. In the generalization
of Galilei's results a new difficulty appears, or rather becomes manifest:
acceleration, and also velocity, are quantities which cannot even be precisely
defined without the transition to the infinitely small. This difficulty was
 removed by Newton.\par
From a solution of Kepler's differential equations it is, of course, possible
to calculate the acceleration at each point. The acceleration will be found to
be always directed towards the sun and to be proportional to 1/r$^2$. This is
the most elegant way to derive the law of gravitation. It was done so by
Newton. Whether Newton considered first the easier derivation from Kepler's
third law, is irrelevant. Given the connection between force and acceleration, 
things are so intimately connected, that one method is inconceivable without
the other.\\[2.0ex]
{\bf Acknowledgements. }\\[3.0ex]
It is a pleasure to acknowledge conversations about Kepler's work with
Profs. S. Brandt and H.D. Dahmen.

\end{document}